\documentclass[pra,a4paper,twocolumn,superscriptaddress,longbibliography]{revtex4-2}
\usepackage{amssymb}
\usepackage{amsmath}
\usepackage{amsfonts}
\usepackage{graphicx}
\usepackage{bm}
\usepackage{xcolor}
\usepackage{multirow}
\usepackage{float}
\usepackage{comment}
\usepackage{ulem}
\usepackage{relsize}
\usepackage{cmap}
\usepackage{bbold}
\usepackage{physics}
\usepackage{setspace}

\usepackage[colorlinks,citecolor=blue,linkcolor=red,urlcolor=blue]
{hyperref}

\DeclareMathOperator{\sgn}{sgn}

\DeclareMathOperator{\im}{Im}
\DeclareMathOperator{\re}{Re}

\begin{document}
	
	\title{Attenuation of flexural 
 phonons in free-standing crystalline two-dimensional materials}
		
\author{A. D. Kokovin}	

\affiliation{Moscow Institute for Physics and Technology, 141700 Moscow, Russia}

\affiliation{\hbox{L.~D.~Landau Institute for Theoretical Physics, acad. Semenova av. 1-a, 142432 Chernogolovka, Russia}}

\author{I. S. Burmistrov}	

\affiliation{\hbox{L.~D.~Landau Institute for Theoretical Physics, acad. Semenova av. 1-a, 142432 Chernogolovka, Russia}}

\affiliation{Laboratory for Condensed Matter Physics, National Research University Higher School of Economics, 101000 Moscow, Russia}

\date{\today} 
	
\begin{abstract}
We develop the theory for dynamics of the out-of-plane deformations in flexible two-dimensional materials. We focus on study of attenuation of flexural phonons in free-standing crystalline membranes. We demonstrate that the dynamical renormalization 
does not involve the ultraviolet divergent logarithmic contributions  contrary to the static ones. This fact allows us to find the scaling form of the attenuation, determine its small and large frequency asymptotes, and to derive the exact expression for the dynamical exponent  of flexural phonons in the long wave limit: $\textsf{z}{=}2{-}\eta/2$. Here $\eta$ is the universal exponent controlling the static renormalization of the bending rigidity.  
 Also we determine the dynamical exponent for the long-wave in-plane phonons: $\textsf{z}^\prime{=}(2{-}\eta)/(1{-}\eta/2)$. 
We discuss implication of our results to experiments on  phonon spectra in graphene and   dynamics of graphene-based
 nanomechanical resonators. 
\end{abstract}

\maketitle
	
\section{Introduction}

Following the discovery of graphene~\cite{Novoselov2004,Novoselov2005,Zhang2005} and other atomically  thin  materials \cite{Novoselov2012}, flexible two-dimensional (2D)  materials \cite{2Dmat} have been attracting a lot of theoretical and experimental interest. These materials,
the so-called crystalline membranes, have a peculiar elastic properties dubbed as anomalous  elasticity. The latter includes non-trivial scaling  of elastic modules with the system size, crumpling transition with increasing  temperature 
and disorder, nonlinear Hooke's law, 
negative Poisson ratios, etc. \cite{Nelson1987,Aronovitz1988,Paczuski1988,David1988,Aronovitz1989,Guitter1988,Guitter1989,Toner1989,Doussal1992,Morse:1992,Nelson_1991,Radzihovsky1991,Morse:1992b,Bensimon_1992,Radzihovsky1995,Radzihovsky1998}. Currently there is a substantial progress in further theoretical understanding of static properties of crystalline membranes \cite{Kats2014,Gornyi:2015a,Kats2016,Burmistrov2016,Gornyi2016,Kosmrlj2017,Doussal2018,Burmistrov2018a,Burmistrov2018b,Saykin2020,Saykin2020b,Coquand2020,Mauri2020,Mauri2021,LeDoussal2021,Shankar2021,Mauri2022,Metayer2022,Burmistrov2022,Parfenov2022}.    

Contrary to extensive study of thermodynamics of membranes, there are just a few works (at least to our knowledge) devoted to membrane's dynamics. The renormalization group method developed to study the static elastic properties of $D{=}4{-}\epsilon$ dimensional membranes (with $\epsilon{\ll}1$) has been extended to investigate dynamical exponent for out-of-plane and in-plane phonons \cite{Lebedev2012}. The dynamics of 2D membranes has recently been studied within phenomenological Langevin-type approach \cite{Mizuochi2014,Granato2022,Granato2023,Steinbock2023}. Intriguing, the dynamical exponents predicted in both mentioned above approaches differ from each other. To resolve the issue, the microscopic theory for the attenuation of flexural phonons in 2D crystalline materials is needed to be developed.  One more motivation for such a theory comes from recent measurement of the phonon spectrum in graphene by the method of the high resolution electron energy loss spectroscopy \cite{Li2023}.

A detailed theory for the attenuation of flexural phonons 
(due to nonlinear effects induced by coupling between in-plane and
out-of-plane displacements)
is not only of an academic interest. Graphene and other 2D crystalline materials are intensively explored as nanoelectromechanical systems with relatively high quality factors  \cite{Miao2014,Leeuwen2014} (see Refs. \cite{Steeneken2021,Ferrari2023} for a review). Also a real-time height dynamics of a free-standing graphene membrane has recently been monitored \cite{Ackerman2016}. Althought there could be many microscopic sources for damping of graphene mechanical nanoresonators \cite{Seoanez2007}, the flexural phonon decay is unavoidable source for intrinsic contribution to damping.

In this paper, we develop the comprehensive theory of the decay time 
($\tau_k$) 
of out-of-plane phonons in free-standing 2D crystalline membranes. We focus on an experimentally relevant temperature range in which flexural phonons can be treated classically, $k_B T{\gg} \hbar \omega_k$. We establish an unexpected result that the decay rate of long wave flexural phonons is independent of temperature and is of the order of the phonon frequency, $1/\tau_k{\sim}\omega_k$. Also we determine exactly the dynamical exponent  for the long-wave flexural phonons: $\omega_k{\sim}k^z$, $\textsf{z}{=}2{-}\eta/2$, cf. Eq. \eqref{eq:z:final}.  Here $k$ stands for the phonon momentum and $\eta$ is the universal exponent controlling the static renormalization of the bending rigidity. We derive similar relation for the spectrum of in-plane phonons with the corresponding dynamical exponent $\textsf{z}^\prime{=}(2{-}\eta)/(1{-}\eta/2)$. 
As application of our results we   compute the time-dependent pair correlation function of membrane's height, cf.~Eq. \eqref{eq:tt-d-hh}.    

\begin{figure}[b]
\centerline{\includegraphics[width=0.44\textwidth]{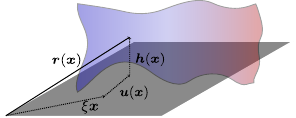}}
\caption{A sketch of rippled membrane with dynamical fluctuations (colored) and reference plane (grey).
}\label{pic: ripples} 
\end{figure}

The outline of the paper is as follows. In Sec. \ref{Sec:Model} we formulate the model of elastic deformations of 2D membrane and announce our main results. 
In Sec. \ref{Sec:Renorm:Stat} we remind a reader the results for the static renormalization of the theory. The computation of the flexural phonon attenuation is presented in Sec. \ref{Sec:Decay}. We explain why there is no effect of dynamics on the crumpling transition in Sec. \ref{Sec:StaticCor}. In Sec. \ref{Sec:Time} we compute the time dependence of pair correlation function of out-of-plane displacement.
We end the paper with discussions and conclusions (Sec. \ref{Sec:DiscConc}). Details of computations are deligated to Appendices. Throughout the paper we use unites with $k_B{=}\hbar{=}1$.

\section{Model and main results\label{Sec:Model}}

The theory  of elasticity of clean 2D crystalline membranes embedded in $d{=}3$ dimensional space is  given by the following free energy
\cite{Nelson1987,Aronovitz1988,Paczuski1988}:
\begin{equation}
\mathcal{F}=  \int d^2 \bm{x} \Bigl ( \frac{\varkappa}{2}(\Delta \bm{r})^2+\mu
u_{\alpha\beta} u_{\alpha\beta}
+\frac{\lambda}{2} u_{\alpha\alpha}^2
\Bigl ) .
\label{eq:FreeEnergy:start}
\end{equation}
Here $\bm{x}$ is the $d{=}2$ coordinate vector of a point on the reference plane while $\bm{r}$ stands for a $d{=}3$ dimensional vector parameterizing a point on the membrane (see Fig.~\ref{pic: ripples}). We introduced the deformation tensor $u_{\alpha\beta}{=}(\partial_\alpha \bm{r} \partial_\beta \bm{r}{-}\delta_{\alpha \beta})/2$,  with $\alpha,\beta{=}x,y$. 
The  bending rigidity is denoted by $\varkappa$ while $\lambda$ and $\mu$ are Lam\'{e} coefficients.

In order to describe the membrane which is not close to the crumpling transition, it is convenient to separate homogeneous stretching ($\xi$) of the membrane, parameterizing the 3D vector $\bm{r}$ as
\begin{equation}
    r_1 = \xi x + u_x, \quad r_2 = \xi y +u_y, \quad r_{3} = h .
\end{equation}
Then the deformation tensor acquires the following form $u_{\alpha\beta}{=} (\xi^2{-}1)\delta_{\alpha\beta}/2{+}\tilde{u}_{\alpha\beta}$, where (no summation over repeating indices is assumed)
\begin{equation}
\tilde{u}_{\alpha\beta}= \frac{1}{2}\Bigl (
\xi_\beta\partial_\alpha u_\beta +\xi_\alpha\partial_\beta u_\alpha+\partial_\alpha h \partial_\beta h + \partial_\alpha \bm{u}\partial_\beta \bm{u} \Bigr ).
\label{eq:u:alpha:beta}
\end{equation}
An inhomogeneous deformation of the membrane is characterized by the $d{=}2$ in-plane displacement vector $\bm{u}{=}\{u_x,u_y\}$ and the scalar out-of-plane deformation $h$.

In order to study dynamics of the in-plane and out-of-plane fluctuations we will work within the path integral formulation in the imaginary time. The partition function is given as
\begin{equation}
    Z = \int\mathcal{D}[h,\bm{u}]
    \exp \Biggl [ - \int\limits_0^{\beta} d\tau \Bigl ( \frac{\rho}{2} 
    \int d^2 \bm{x}\, (\partial_\tau \bm{r})^2 + \mathcal{F} \Bigr )
    \Biggr ] .
    \label{eq:Z:start}
\end{equation}
Here $\beta {=} 1/T$ is the inverse temperature and $\rho$ is the mass density of the membrane. 

Provided the membrane is in the flat phase away from the crumpling transition, it is legitimate \cite{Nelson1987} to omit the term  $\partial_\alpha \bm{u}\partial_\beta \bm{u}$ in Eq. \eqref{eq:u:alpha:beta}. Similarly, one can neglect the contribution from $\bm{u}$ to the bending energy.
Then the free energy  Eq. \eqref{eq:FreeEnergy:start} becomes quadratic in terms of the in-plane displacements. It allows us to integrate over $\bm{u}$ in Eq. \eqref{eq:Z:start} exactly and to derive the effective action for the out-of-plane displacement alone (see details of derivation in Refs. \cite{Gornyi:2015a,Burmistrov2016}),
\begin{equation}
Z = \int \mathcal{D}[h] e^{-S}, \qquad
S=S_0+S_{\rm dyn}, 
\end{equation}
where 
\begin{equation}
S_0 = \frac{\beta}{8}\int d^2\bm{x} \, c_{\alpha\beta} \varepsilon_\alpha\varepsilon_\beta, \quad \varepsilon_\alpha{=}\xi^2{-}1{+}  \sum_{\omega,\bm{k}}
k^2_\alpha |h_{\bm{k},\omega}|^2 .
\label{eq:action:1}
\end{equation}
and
\begin{align}
S_{\rm dyn}= & \frac{1}{2} \sum_{\omega,\bm{k}}\left(\varkappa k^4 + \rho \omega^2 \right)\left| h_{\bm{k},\omega}\right|^2 \notag \\
& +  \frac{Y}{8} 
\sum_{\Omega,\bm{q}{\neq} 0}
 \Biggl | \sum_{\omega,\bm{k}} \frac{[\bm{k}\times{\bm{q}}]^2}{q^2} h_{\bm{k+q},\omega+\Omega}h_{-\bm{k},-\Omega}\Biggr |^2 .
\label{eq:action:2}
\end{align}
Here $c_{\alpha\beta}{=}\lambda {+}2\mu\delta_{\alpha\beta}$ stands for the matrix of elastic stiffness constants and $Y {=} 4\mu (\mu {+} \lambda)/(2\mu {+} \lambda)$ is the Young's modulus. Also we performed the Fourier transform
\begin{equation}
    h(\bm{x},\tau) = \sum\limits_{\omega_n, \bm{k}} h_{\bm{k},\omega_n} e^{i (\bm{k} \bm{x} {-} \omega_n \tau)} ,
    \label{eq:h:FT}
\end{equation}
where $\omega_n {=} 2\pi T n$ are the bosonic Matsubara frequencies. Here and in what follows, we use the short-hand notation $\sum_{\omega_n, \bm{k}} {=} T \sum_{\omega_n}\int {d^2k}/{(2\pi)^2}$. We note that the term $\sum_{\omega,\bm{k}}
k^2_\alpha |h_{\bm{k},\omega}|^2$ in the displacement $\varepsilon_\alpha$ is   responsible for the anomalous Hooke's law. 

Generally,  due to dynamics of the in-plane phonons, the interaction in the second line of Eq. \eqref{eq:action:2}, i.e the Young's modulus $Y$, depends on the transferred frequency $\Omega$, see Ref. \cite{Burmistrov2016}. However, as one can check, the static limit of interaction mediated by the in-plane phonons is enough for our computations (see Appendix \ref{Appendix:1:Inplane} for details).

The quadratic part of action \eqref{eq:action:2} determines the ``bare'' 
Green's function in the Matsubara representation
\begin{equation}
G_{\bm{k}}(i\omega) = \frac{1}{\rho \omega^2 + \varkappa k^4} .
\label{eq:nonint:G}
\end{equation}
The corresponding retarded and advanced Green's functions are given as 
\begin{equation}
G_{\bm{k}}^{R/A}(\omega) = - \frac{1}{\rho(\omega\pm i0^+)^2 - \varkappa k^4}  .
\label{eq:nonint:GRA}
\end{equation}
Using  Eq.~\eqref{eq:nonint:GRA} one can extract the spectrum of non-interacting flexural phonons: 
\begin{equation}
\omega_k^{(0)} = D k^2, \qquad D= \sqrt{\varkappa/\rho} .
\label{eq:bare:spectrum:f}
\end{equation}

Since the theory \eqref{eq:action:2} is interacting, the exact Green's function is related with the bare one by the Dyson equation
\begin{equation}
\begin{split}
\mathcal{G}^{-1}_{\bm{k}}(i\omega) & =
G^{-1}_{\bm{k}}(i\omega)] -\Sigma_{\bm{k}}(i\omega) , \\
[\mathcal{G}^{R/A}_{\bm{k}}(\omega)]^{-1} & =
[G^{R/A}_{\bm{k}}(\omega)]^{-1} -\Sigma_{\bm{k}}^{R/A}(\omega) .
\end{split}
\label{eq:exact:G}
\end{equation}

In this paper our aim is to compute the frequency dependence of the retarded self energy $\Sigma_{\bm{k}}^R(\omega)$. As usual, it is related with $\Sigma_{\bm{k}}(i\omega)$ by analytic continuation $i\omega{\to}\omega{+}i0^+$. The static self energy $\Sigma_{\bm{k}}^R(0)$ was studied in many works before. It is well established that the perturbation theory in powers of interaction produces ultra-violet logarithmic divergences that can be summed up by means of the renormalization group (RG). The emergent ultra-violet scale is the so-called inverse Ginzburg length, $q_*{=}\sqrt{3Y T/(32\pi \varkappa^2)}$. Such RG-improved perturbation theory results in a power law renormalization of the bending rigidity and Young's modulus \cite{David1988,Aronovitz1988}
\begin{equation}
\varkappa(k) {=}\varkappa (q_*/k)^\eta,
\quad Y(k) {=} Y (q_*/k)^{2{-}2\eta}, \quad k\ll q_* ,
\label{eq:ren:12}
\end{equation}
where the universal exponent $\eta{\simeq}0.795{\pm}0.01$ is determined numerically \cite{Troster2013}. 

\begin{figure}[t]
    \centering
    \centerline{\includegraphics[width = 0.85\columnwidth]{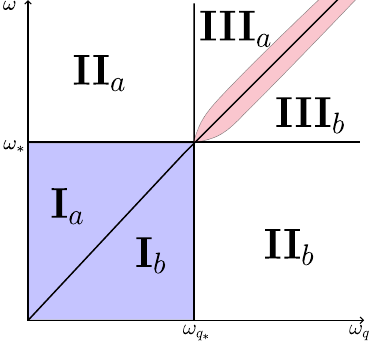}}
    \caption{A sketch of the range of momenta and frequencies (universal regime) considered in the paper, see Eq. \protect\eqref{eq:range}.
    }
    \label{fig:UniversalRegime}
\end{figure}

It is convenient to introduce the frequency scale corresponding to the Ginzburg length, $\omega_*{=}D q_*^2$.
 Introducing the dimensionless parameter characterizing the strength of quantum effects for membrane, $g{=} 21 Y/(128\pi\sqrt{\rho \varkappa^3})$ \cite{Kats2014,Kats2016,Burmistrov2016}, we find that $\omega_*{=}(4/7) g T$. In what follows we will assume that $g{\ll} 1$ (e.g. for graphene $g{\approx}0.05$). Also we will consider the following range of momenta and frequencies, see Fig. \ref{fig:UniversalRegime},
\begin{equation}
 k\ll q_*, \qquad |\omega|\ll\omega_*\ll T . 
 \label{eq:range}
\end{equation}

Below we will call the regime \eqref{eq:range} as the universal regime.

\begin{figure}
    \centering
    \centerline{\includegraphics[width = 0.85\columnwidth]{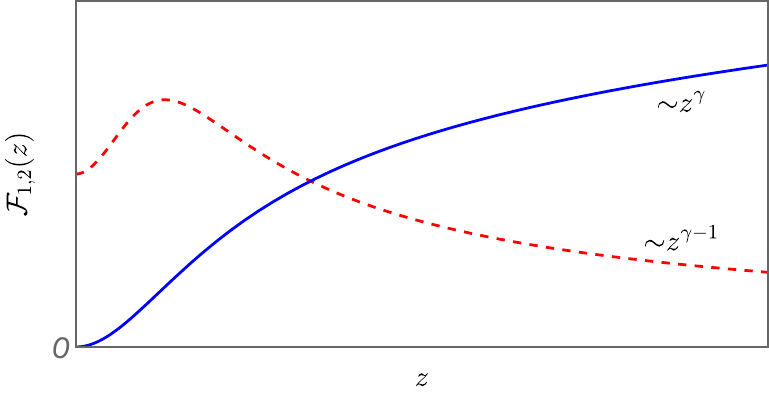}}
    \caption{The sketch of the behavior of the functions $\mathcal{F}_{1}(z)$ (blue solid curve) and $\mathcal{F}_{2}(z)$ (red dashed curve). The exponent $\gamma$ is defined in \eqref{eq:def:gamma}.}
    \label{fig:F1&F2}
\end{figure}

We demonstrate below that the retarded self energy in the range \eqref{eq:range} can be written in the following scaling form 
 \begin{equation}
 \begin{split}
 \Re \Sigma^R_{\bm{k}}(\omega)-\Sigma^R_{\bm{k}}(0)& =
\rho \omega_k^2 \mathcal{F}_1(\omega/\omega_k),
\\ 
\Im\Sigma^R_{\bm{k}}(\omega) & =\rho\omega\omega_k \mathcal{F}_2(\omega/\omega_k) .
\end{split}
\label{eq:Green:main:0}
\end{equation}
This is the main result of our work.
Here we introduce 
\begin{equation}
\omega_k = D k^2 (k/q_*)^{{-}\eta/2} \sim k^{\textsf{z}}, \quad \textsf{z}=2-\eta/2 ,
\label{eq:exact:ff}
\end{equation}
that is upto an unknown numerical factor describes the exact spectrum of a flexural phonon. The scaling functions $\mathcal{F}_1(z)$ and $\mathcal{F}_2(z)$  are even functions of their argument, satisfy the normalization condition $\mathcal{F}_1(0){=}0$, and obey  Kramers-Kronig-type relations,
\begin{equation}
\begin{split}
\mathcal{F}_1(z) &=  {\rm p.v.} \int_{-\infty}^\infty\frac{dx}{\pi} \frac{z \mathcal{F}_2(x)}{x-z}, \\  \mathcal{F}_2(z) &=  {\rm p.v.} \int_{-\infty}^\infty\frac{dx}{\pi} \frac{ \mathcal{F}_1(x)}{z(z-x)} .
\end{split}
\label{eq:KK}
\end{equation}
The qualitative behavior of functions $\mathcal{F}_1(z)$ and $\mathcal{F}_2(z)$ is shown in Fig. \ref{fig:F1&F2}.

The relations \eqref{eq:Green:main:0} implies the following scaling form of the exact retarded Green's function
\begin{gather}
\mathcal{G}^{R}_{\bm{k}}(\omega){=}
{-} \frac{1}{\rho}\left [\omega^2 {-}\omega_k^2 \Bigl [1{-}\mathcal{F}_1\Bigl (\frac{\omega}{\omega_k}\Bigr)\Bigr ]{+}i \omega \omega_k \mathcal{F}_2\Bigl (\frac{\omega}{\omega_k}\Bigr )\right ]^{{-}1} .
\label{eq:Green:main}
\end{gather}

In the next two sections, Secs. \ref{Sec:Renorm:Stat} and \ref{Sec:Decay}, we will explain how the results \eqref{eq:Green:main:0} can be derived and present asymptotic expressions for the functions $\mathcal{F}_{1,2}$. Physical implications of the result 
\eqref{eq:Green:main} are discussed
in Secs. \ref{Sec:StaticCor} and \ref{Sec:Time}.


\section{Static renormalization\label{Sec:Renorm:Stat}}

The theory of static out-of-plane displacements was extensively explored previously (see Ref.  \cite{Doussal2018} for a review). In this section, we remind a reader how these results, in particular, Eq. \eqref{eq:ren:12}, can be derived within frequency dependent Green's functions.

\begin{figure}[t]
\centering
\centerline{\includegraphics[width=0.85\columnwidth]{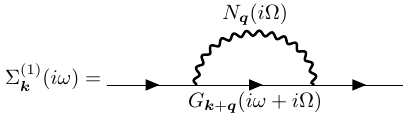}}   
\centerline{\includegraphics[width=0.85\columnwidth]{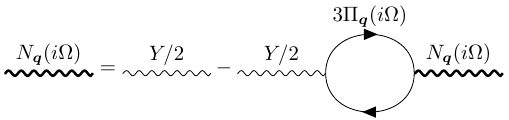}}
    \caption{(Top) SCSA-like contribution to the self energy correction, (Bottom) RPA-like screened interaction.}
    \label{pic:SelfEnergy}
\end{figure}

Let us start from the self energy contribution shown in Fig.~\ref{pic:SelfEnergy},
\begin{equation}
\Sigma^{(1)}_{\bm{k}}(i\omega)=
-2 \sum_{\Omega,\bm{q}}
\frac{[\bm{k}\times \bm{q}]^4}{q^4} N_{\bm{q}}(i\Omega) 
G_{\bm{k}+\bm{q}}(i\omega+i\Omega) .
\label{eq:Sigma:Im:Pert:0:M}
\end{equation}
It is the first order correction to the self energy in the dynamical RPA-type screened interaction (see Fig.~\ref{pic:SelfEnergy}),
\begin{gather} 
N_{\bm{q}}(i\Omega) = \frac{Y/2}{1+3Y\Pi^{(0)}_{\bm{q}}(i\Omega)/2} .
\label{eq:N:screened}
\end{gather}
Here the `bare' polarization operator is given as
\begin{equation}
\Pi^{(0)}_{\bm{q}}(i\Omega) = 
\frac{1}{3} \sum_{\omega,\bm{k}} 
\frac{[\bm{k}\times \bm{q}]^4}{q^4}
G_{\bm{k}}(i\omega)
G_{\bm{k}+\bm{q}}(i\omega+i\Omega)  .
\label{eq:Pol:Im:Pert:0:M}
\end{equation}
We emphasize that RPA-type screening is crucial in the region $q{\ll}q_*$ since 
$Y\Pi^{(0)}_{\bm{q}}(0){\sim} (q_*/q)^2{\gg} 1$.
Making the analytic continuation in Eq.~\eqref{eq:Sigma:Im:Pert:0:M} to the real frequencies, $i\omega{\to}\omega{+}i0$, we find
\begin{gather}
\Sigma^{(1),R}_{\bm{k}}(\omega) {=} 
 {-}\int \!\frac{d\Omega}{\pi}\!\!
\int \!\frac{d^2\bm{q}}{(2\pi)^2}
\frac{[\bm{k}{\times}\bm{q}]^4}{q^4} \Biggl[ \coth \frac{\Omega}{2T}
\Im N^R_{\bm{q}}(\Omega) \notag \\ \times G^R_{\bm{k}{+}\bm{q}}(\omega{+}\Omega) 
{+} \coth \frac{\omega{+}\Omega}{2T}N^A_{\bm{q}}(\Omega) \Im G^R_{\bm{k}{+}\bm{q}}(\omega{+}\Omega) 
\Biggr ] .
\label{eq:Sigma:Pert:0}
\end{gather}
Here we introduced retarded dynamically screened interaction, $N^{R}_{\bm{q}}(\Omega) {=} (Y/2)/[1{+}3Y\Pi^{(0),R}_{\bm{q}}(\Omega)/2]$, where 
\begin{gather}
\Pi^{(0),R}_{\bm{q}}(\Omega) {=} 
 \int \!\frac{d\omega}{2\pi}
\!\!\int \! \frac{d^2\bm{k}}{(2\pi)^2}
\frac{[\bm{k}\times \bm{q}]^4}{3q^4}\Biggl \{
\coth \frac{\omega}{2T} \im G^{R}_{\bm{k}}(\omega)  \notag \\ \times G^{R}_{\bm{k}{+}\bm{q}}(\omega{+}\Omega) 
{+}  \coth \frac{\omega{+}\Omega}{2T} G^{A}_{\bm{k}}(\omega) \im G^{R}_{\bm{k}{+}\bm{q}}(\omega+\Omega) 
\Biggr \}
 .\label{eq:pol:R}
\end{gather}
is the retarded polarization operator corresponding to the Matsubara one, cf. Eq. \eqref{eq:Pol:Im:Pert:0:M}.
We note that $N^{A}_{\bm{q}}(\Omega)$ can be obtained from $N^{R}_{\bm{q}}(\Omega)$ by complex conjugation.

Setting in Eq. \eqref{eq:Sigma:Im:Pert:0:M} the frequency $\omega$ to zero, we obtain
\begin{gather}
\re \Sigma^{(1),R}_{\bm{k}}(0) = 
 -\int \frac{d\Omega}{\pi} \coth\frac{\Omega}{2T}
\int \frac{d^2\bm{q}}{(2\pi)^2}
\frac{[\bm{k}\times \bm{q}]^4}{q^4}
\notag \\ 
\times \im \Bigl[ N^R_{\bm{q}}(\Omega) G^R_{\bm{k}+\bm{q}}(\Omega) \Bigr ]
 .
\label{eq:Sigma:Pert:0:1}
\end{gather}
In the classical regime, $T{\gg}|\Omega|$ we can use the following approximation, $\coth(\Omega/2T) {\sim} 2T/\Omega$. Then, we perform the integral over $\Omega$ in Eq. \eqref{eq:Sigma:Pert:0:1} with the help of Kramers--Kronig relation. Eventually, we find
\begin{equation}
\re \Sigma^{(1),R}_{\bm{k}}(0)=
-2 T \int \frac{d^2\bm{q}}{(2\pi)^2}
\frac{[\bm{k}\times \bm{q}]^4}{q^4} 
 N^R_{\bm{q}}(0) G^R_{\bm{k}+\bm{q}}(0).
 \label{eq:Sigma:Pert:0:2}
\end{equation}
Comparison of Eq. \eqref{eq:Sigma:Pert:0:2} with Eq. 
\eqref{eq:Sigma:Im:Pert:0:M} shows that Eq. \eqref{eq:Sigma:Pert:0:2} fully reproduces the result of static treatment. 


A similar procedure can be performed for all other diagrams as well. For example, for 
the diagram shown in Fig. \ref{pic:Non-SCSA} we find (see Appendix \ref{Appendix:2:Non-SCSA-diag:static}):
\begin{gather}
\re\Sigma_{\bm{k}}^{(2), R}(0){=}{-}4 T^2 \sum_{\bm{q}, \bm{Q}} \frac{[(\bm{k}{+}\bm{Q}) {\times} \bm{q}]^2}{q^2}  \frac{[(\bm{k}{+}\bm{q}) {\times} \bm{Q}]^2}{Q^2} 
\frac{[\bm{k} {\times} \bm{q}]^2}{q^2}   \notag \\ 
\times
\frac{[\bm{k} {\times} \bm{Q}]^2}{Q^2}
 G^R_{\bm{k}{+}\bm{q}}(0) G^R_{\bm{k}{+}\bm{Q}}(0)G^R_{\bm{k}{+}\bm{q}{+}\bm{Q}}(0) N_{\bm{q}}^{R}(0)N_{\bm{Q}}^{R}(0) .
 \label{eq:diag:Fig3:static}
\end{gather}

The analysis above can be extended to any self energy diagram with zero external frequency. Indeed, only the static Green's function and static screened interaction contribute to the zero-frequency self energy corrections in the classical regime, $\omega{\ll}T$. 

As we discussed above, 
the diagrams for static self energy are logarithmically divergent and $q_*$ serves as the ultra-violet cut-off. Therefore, it is worthwhile, at first, to sum up all contributions to $\Sigma_{\bm{k}}^R(0)$, and only then to develop perturbation theory for $\im \Sigma_{\bm{k}}^R(\omega)$ (see discussion of similar approach in Ref. \cite{Lebedev2012}).  
This idea implies that new `bare' Green's function for such `dynamical' perturbation theory reads
\begin{gather}
\mathcal{G}^{(0)}_{\bm{k}}(i\omega) = \frac{1}{\rho \omega^2{+}\varkappa k^4 {-}\Sigma^R_{\bm{k}}(0)} \equiv \frac{1}{\rho (\omega^2{+}\omega_k^2)} ,
\label{eq:GreenG:bare:2}
\end{gather}
where $\omega_k$ is given by Eq. \eqref{eq:exact:ff}. We note that the perturbation theory for $\Im\Sigma_{\bm{k}}(\omega)$ consists of the same diagrams as the one for the full self energy but, additionally, a number of diagrams to avoid double counting is needed  to be considered. We discuss this issue in detail in Appendix \ref{Appendix:3:Dyn:Pert}. Although, due to counter-terms such a diagrammatic technique is not convenient beyond the lowest order in interaction, nevertheless, it has an important advantage: as we will demonstrate below the diagrams computed with the help of the Green's function with the statically renormalized phonon spectrum, Eq. \eqref{eq:GreenG:bare:2}, are convergent in the ultra-violet. 

\begin{figure}[t]
    \centering
\centerline{\includegraphics[width =0.95\columnwidth]{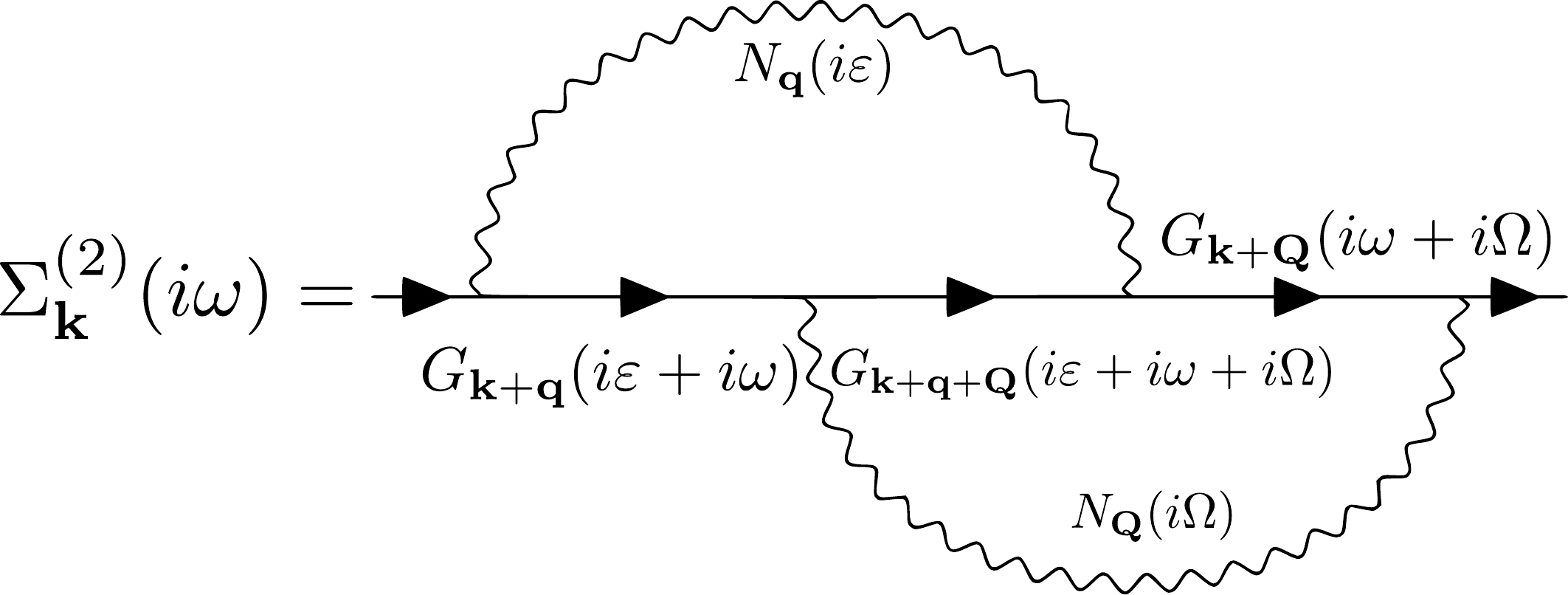}}
    \caption{Non-SCSA-like contribution to the self energy correction.}
    \label{pic:Non-SCSA}
\end{figure}

\section{Interaction-induced flexural phonon decay\label{Sec:Decay}}

\begin{figure}[t]
    \centering
    \centerline{\includegraphics[width = 0.75\columnwidth]{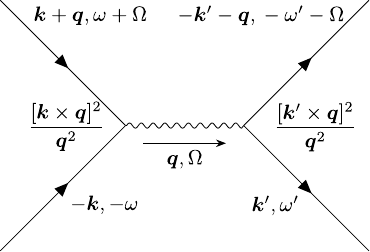}}
    \caption{Diagram illustrating a four-phonon process.}
    \label{pic:4phonons}
\end{figure}

Now we are ready to compute the imaginary part of the self energy that determines the decay of flexural phonons. The source of decay is the four-phonon processes, see Fig. \ref{pic:4phonons}, due to the interaction term in the second line of Eq.~\eqref{eq:action:2}.

We start from the diagram shown in Fig. \ref{pic:SelfEnergy}. Taking the imaginary part of the expression \eqref{eq:Sigma:Pert:0:1}, we find the following result in the universal regime (regions I$_a$ and I$_{b}$ in Fig.~\ref{fig:UniversalRegime}), 
\begin{gather}
\Im \Sigma^{(1),R}_{\bm{k}}(\omega) {=} 
 {-} \frac{2T\omega}{3} \int \frac{d\Omega}{\pi}\!
\int \frac{d^2\bm{q}}{(2\pi)^2}
\frac{[\bm{k}{\times} \bm{q}]^4}{q^4}
\frac{\Im \Pi^{(0),R}_{\bm{q}}(\Omega)}{|\Pi^{(0),R}_{\bm{q}}(\Omega)|^2}
\notag \\
\times  
\frac{\Im \mathcal{G}^{(0),R}_{\bm{k}+\bm{q}}(\omega+\Omega)}{\Omega(\omega+\Omega)} .
\label{eq:ImSigma:Pert:1:R}
\end{gather}
Here we substituted $G$ by $\mathcal{G}^{(0)}$. Also the polarization operator $\Pi^R_{\bm{q}}(\Omega)$ is given by Eq. \eqref{eq:pol:R} with the Green's function $G$ substituted by $\mathcal{G}^{(0)}$. Before analysing 
the correction \eqref{eq:ImSigma:Pert:1:R}, we discuss the frequency dependence of the polarization operator.

\subsection{Polarization operator}

Taking the imaginary part of the right hand side of Eq. \eqref{eq:pol:R}, we obtain the following expression in the universal regime
\begin{gather}
\im \Pi^{(0),R}_{\bm{q}}(\Omega) = 
\frac{2 T \Omega}{3}  \int \frac{d\omega}{2\pi}
\int \frac{d^2\bm{k}}{(2\pi)^2}
\frac{[\bm{k}{\times} \bm{q}]^4}{q^4} \frac{\im \mathcal{G}^{(0),R}_{\bm{k}}(\omega)}{\omega}
\notag \\
\times \frac{\im \mathcal{G}^{(0),R}_{\bm{k}+\bm{q}}(\omega+\Omega)}{\omega+\Omega} \ .
\label{eq:pol:R:im}
\end{gather}
Neglecting the external frequency $\Omega$ under the integral signs in Eq. \eqref{eq:pol:R:im}, we find the following asymptotic behavior at $\Omega{\to} 0$, 
\begin{equation}
\im \Pi^{(0),R}_{\bm{q}}(\Omega) \propto \frac{T}{\varkappa^2 q^{2-2\eta} q_*^{2\eta}} \frac{\Omega}{\omega_q}, \quad |\Omega|\ll\omega_q .  
\label{eq:ImP:small:as}
\end{equation}
In the opposite case of high frequencies, we obtain 
\begin{equation}
\im \Pi^{(0),R}_{\bm{q}}(\Omega) \propto \frac{T}{\varkappa^2 q^{2-2\eta} q_*^{2\eta}} \left (\frac{\Omega}{\omega_q}\right )^{-\gamma}, \quad |\Omega|\gg\omega_q .  
\label{eq:ImP:high:as}
\end{equation}
where we introduced the exponent 
\begin{equation}
    \gamma = \frac{1-\eta}{1-\eta/4} \simeq 0.256.
    \label{eq:def:gamma}
\end{equation}
The detailed derivation of the above asymptotic results is given in Appendix \ref{Appendix:4:PolOper:0}. 

Equations \eqref{eq:ImP:small:as} and \eqref{eq:ImP:high:as} together with analytic properties suggest the following form of the polarization operator
\begin{equation}
    \Pi^{(0),R}_{\bm{q}}(\Omega) = \frac{A_\eta T q_*^{-2\eta}}{\varkappa^2 q^{2-2\eta}} \left [\mathcal{P}_1^{(0)}\left (\frac{\Omega}{\omega_q}\right ) + i \mathcal{P}_2^{(0)}\left (\frac{\Omega}{\omega_q}\right) \right ] .
    \label{eq:P000}
\end{equation}
Here we introduce numerical factor \cite{Gornyi:2015a}
\begin{equation}
A_\eta = \frac{\Gamma(1{+}\eta/2)\Gamma(1{-}\eta)}{2^{5{+}\eta}\sqrt{\pi}\Gamma^2(2{-}\eta/2)\Gamma((3{+}\eta)/2)}
\end{equation}
to ensure the normalization condition, $\mathcal{P}_1^{(0)}(0){=}1$. As it follows from Eqs. \eqref{eq:ImP:small:as} and \eqref{eq:ImP:high:as}, the odd function $\mathcal{P}_2^{(0)}(z)$ has the following asymptotic behavior
\begin{equation}
  \mathcal{P}_2^{(0)}(z) \propto \begin{cases}
      z, & \quad |z|\ll 1 , \\
      \sgn z \, |z|^{-\gamma}, & \quad |z| \gg 1 .
  \end{cases}  
  \label{eq:ppp:01}
\end{equation}
Thus the function $\mathcal{P}^{(0)}_{2}(z)$ has extrema at $|z|{\sim} 1$.

In order to determine the asymptotic behavior of the real part of the polarization operator at finite frequency, i.e. the function $\mathcal{P}_1^{(0)}(z)$, we use the Kramers-Kronig relation:
\begin{gather}
\mathcal{P}^{(0)}_1(z) =1+\frac{2 z^2}{\pi} {\rm p.v.} \int\limits_0^\infty \frac{d y}{y} \frac{\mathcal{P}^{(0)}_2(y)}{y^2-z^2}  .
\label{eq:PP:KK}
\end{gather}
Neglecting $z$ under the integral sign in Eq. \eqref{eq:PP:KK}, we find 
\begin{equation}
\mathcal{P}^{(0)}_1(z) \simeq  1 +\frac{z^2}{\pi} \int\limits_0^\infty \frac{dy}{y} \frac{d}{dy}  \frac{\mathcal{P}_2^{(0)}(y)}{y} , \quad |z|\ll 1 .
\label{eq:ReP:small:as}
\end{equation}
At large magnitudes of the argument, we obtain (see Appendix \ref{Appendix:4:PolOper:0})
\begin{equation}
    \mathcal{P}^{(0)}_1(z) \propto z^{-\gamma}, \quad |z| \gg 1.
\label{eq:ReP:high:as}
\end{equation}

We note that at high frequencies, $\Omega{\gg}\omega_q$, the polarization operator is independent of the momentum, $\Pi^{(0),R}_{\bm{q}}(\Omega){\sim}\Omega^{{-}\gamma}$. This fact can be naturally understood. One needs to take the static polarization operator and substitute the momentum $q_* (\Omega/\omega_*)^{1/(2-\eta/2)}$ instead of $q$. The former momentum corresponds to the mass shell condition, $\Omega{=}\omega_q$. We note that such situation 
 is consistent with 
the dynamical exponent $\textsf{z}{=}2{-}\eta/2$ (see more detail in Sec. \ref{Sec:Time}).

\subsection{Result for the first order self energy correction}

Now we turn back to Eq. \eqref{eq:ImSigma:Pert:1:R}. With known asymptotic behavior of the polarization operator $\Pi^{(0),R}_{\bm{q}}(\Omega)$, we are able to show (see Appendix \ref{Appendix:4:Sigma:0}) that 
\begin{equation}
    \Im \Sigma^{(1),R}_{\bm{k}}(\omega) = \rho \omega \omega_k \mathcal{F}_2^{(1)}\left (\frac{\omega}{\omega_k}\right ) ,
    \label{eq:ImS:00}
\end{equation}
where the even function $\mathcal{F}_2^{(1)}(z)$ has the following asymptotic behavior 
\begin{equation}
\begin{split}
   \mathcal{F}_2^{(1)}(z) - 
    \mathcal{F}_2^{(1)}(0) \propto z^2,  \quad |z|\ll 1 , \\
    \mathcal{F}_2^{(1)}(z) \propto |z|^{\gamma-1}, \quad |z|\gg 1  .
    \end{split}
    \label{eq:F21:00:as}
\end{equation}
We emphasize that $\Im \Sigma^{(1),R}_{\bm{k}}(\omega)$ is given by the ultra-violet convergent integrals and, consequently, it does not involve the frequency scale $\omega_*$. 

The real part of the self energy correction can be parametrized in a similar way as the imaginary one,
\begin{equation}
\re \Sigma^{(1),R}_{\bm{k}}(\omega) - \re \Sigma^{(1),R}_{\bm{k}}(0) = 
\rho \omega_k^2 \mathcal{F}_1^{(1)}\left (\frac{\omega}{\omega_k}\right ) .
\label{eq:ReS:00}
\end{equation}
Here the even function $\mathcal{F}_1^{(1)}(z)$ is related with $\mathcal{F}_2^{(1)}(z)$ by Kramers-Kronig-type relation
\begin{equation}
\mathcal{F}_1^{(1)}(z)=  {\rm p.v.} \int\limits_{-\infty}^\infty\frac{ dx}{\pi} \frac{z \mathcal{F}_2^{(1)}(x)}{x-z} .
\label{eq:KK:0}
\end{equation}
Using asympotics of $\mathcal{F}_2^{(1)}(z)$ we find the following behavior of $\mathcal{F}_1^{(1)}(z)$ at small and large arguments (see Appendix~\ref{Appendix:4:Sigma:0}),
\begin{equation}
   \mathcal{F}_1^{(1)}(z) \propto 
   \begin{cases}
 z^2, & \quad |z|\ll 1 ,
 \\
 |z|^{\gamma}, &  \quad |z|\gg 1 .
 \end{cases}
 \label{eq:F11:00:as}
\end{equation}
We emphasize that the frequency integral in the Kramers-Kronig relation \eqref{eq:KK:0} is convergent in the ultra-violet such that there is no need in $\omega_*$ as ultra-violet cut-off
for computation of the function $\mathcal{F}_1^{(1)}(z)$.

\subsection{Analysis of higher order diagrams}

\begin{figure}
    \centering
    \includegraphics[width = 0.85\columnwidth]{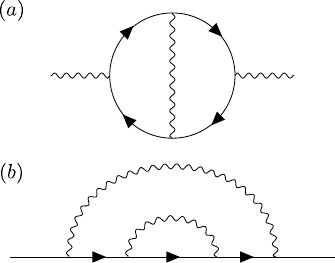}
    \caption{(a) Correction to the polarization operator, (b) Correction to the self energy in the next order in the interaction}
    \label{pic:Higher_Order}
\end{figure}

In general, there is no reason to limit computation of the dynamical self energy just by the lowest order diagram shown in Fig. \ref{pic:SelfEnergy}. Moreover, even for that diagram, the polarization operator should be computed in the next orders in the interaction. We show  examples of higher order diagrams in Fig. \ref{pic:Higher_Order}.  Although the analytical computation of all necessary diagrams is hopeless, we can compute asymptotic behavior of both the exact polarization operator and the exact self energy. 

Assuming that frequency behavior of the self energy is the same as given by Eqs. \eqref{eq:F21:00:as} and \eqref{eq:F11:00:as}, one can check that the exact polarization operator retains the same scaling form as in Eq. \eqref{eq:P000} (see Appendix \ref{Appendix:6:PolOper:Exact} for details). So, we find that the exact polarization operator can be written as 
\begin{equation}
    \Pi^{R}_{\bm{q}}(\Omega) = \frac{T}{\varkappa^2 q^{2-2\eta}q_*^{2\eta}} \left [\mathcal{P}_1\left (\frac{\Omega}{\omega_q}\right ) + i \mathcal{P}_2\left (\frac{\Omega}{\omega_q}\right ) \right ] ,
    \label{eq:PolOp:main}
\end{equation}
where $\mathcal{P}_1(z)$ and $\mathcal{P}_2(z)$ are even and odd functions of $z$, respectively. They have the following asymptotic behavior, 
\begin{equation}
  \mathcal{P}_1(z){=}\mathcal{P}_1(0) (1{+} B_1^{(0)} z^2), \quad \mathcal{P}_2(z) {=}B_2^{(0)} z, \quad |z|{\ll} 1 , 
  \label{eq:PolOp:main:A1}
\end{equation}
and 
\begin{equation}
   \mathcal{P}_{1}(z) {=} B_1^{(\infty)} |z|^{{-}\gamma}, \quad \mathcal{P}_{2}(z) {=} B_2^{(\infty)} \sgn z \,  |z|^{{-}\gamma}, \quad |z|{\gg} 1 ,
   \label{eq:PolOp:main:A2}
\end{equation}
where $B_{1,2}^{(0,\infty)}$ are numerical coefficients. We note that we do not normalize $\mathcal{P}_{1}(0)$ to be equal to unity. In virtue of the Kramers-Kronig relations we find relations the numerical coefficients intriduced above have to satisfy,
\begin{equation}
\mathcal{P}_1(0) {=} \frac{2}{\pi}\int\limits_0^\infty \frac{dx}{x} \mathcal{P}_2(x), \quad B_1^{(0)} {=} \frac{\int_0^\infty dx \,\mathcal{P}_2^{\prime\prime}(x)/x}{\int_0^\infty dx \,\mathcal{P}_2(x)/(2x)} 
\end{equation}
and 
\begin{equation}
B_1^{(\infty)} = - B_2^{(\infty)} \Phi_\gamma , \quad
    \Phi_\gamma = \int\limits_0^\infty \frac{dt}{\pi t} \Bigl [(1+t)^\gamma-|1-t|^\gamma\Bigr ] .
    \label{eq:Phi_gamma}
\end{equation}
 
Now we can use the results \eqref{eq:PolOp:main} - \eqref{eq:PolOp:main:A2} in order to compute higher order diagrams for the self energy whose examples are shown in Fig. \ref{pic:Higher_Order}. Then for the exact Green's function we reproduce the result \eqref{eq:Green:main} (see Appendix \ref{Appendix:7:Sigma:Exact}). The functions $\mathcal{F}_{1,2}(z)$ have the following asymptotics
\begin{equation}
\mathcal{F}_{1}(z) = C_1^{(0)} z^2, \quad
    \mathcal{F}_{2}(z) = \mathcal{F}_{2}(0)(1+C_2^{(0)}z^2), \quad |z|\ll 1 
    \label{eq:F21:zz}
\end{equation}
and 
\begin{equation}
  \mathcal{F}_{1}(z) = C^{(\infty)}_1 |z|^{\gamma} , \quad \mathcal{F}_{2}(z) = C_2^{(\infty)} |z|^{\gamma-1}, \quad |z|\gg 1 ,
  \label{eq:F12:zz}
\end{equation}
where $C_{1,2}^{(0,\infty)}$ are numerical coefficients which satisfy the following relations 
\begin{gather}
  C_1^{(0)} = \frac{2}{\pi}
    \int_0^\infty \frac{dx}{x}\mathcal{F}^\prime_2(x), 
    \quad
    C_1^{(\infty)} = C_2^{(\infty)} \Phi_\gamma .
    \label{eq:Phi_gamma:2}
\end{gather}

We note that the numerical coefficients introduced above for the asymptotic expressions of the exact polarization operator and the self energy can be found within $1/d_c$ expansion for 2D membrane embedded into $d_c+2$ dimensional space \cite{David1988}. We present the results of such calculations in Appendix \ref{Appendix:8:1dc}.

So far we analyze the exact self energy in the universal regime (regions I$_a$ and I$_b$ in Fig. \ref{fig:UniversalRegime}). The behavior of the self energy beyond the universal regime is controlled by the lowest order diagrams and discussed in Appendix \ref{App:I}.

\subsection{Attenuation of flexural phonons}

 The above results proves the form \eqref{eq:Green:main} of the exact Green's function and provides asymtotic expressions for the functions $\mathcal{F}_{1,2}$. 
The exact Green's function in the form of \eqref{eq:Green:main} 
 implies that the spectrum of flexural phonons at $k{\ll}q_*$ is given as 
$\omega {=} s \omega_k $ where a complex number $s$ solves the following equation
\begin{equation}
    s^2-1+\mathcal{F}_1(s)+i s \mathcal{F}_2(s) = 0  .
\end{equation}
The solution of this equation is a complex number $s$ with, generically, $|s|{\sim} 1$. It implies that the imaginary part of the flexural phonon's spectrum $\im s \, \omega_k$ is of the same order as its real part, $\re s \, \omega_k$. In particular, if one defines the decay rate  $1/\tau_k{=}\im\Sigma_{\bm{k}}^R(\omega_k)/(\rho \omega_k)$ then one finds $\omega_k \tau_k{\sim} 1$. This poses several questions: (i) why we do not see implications of such a short decay time in the theory of anomalous elasticity? (ii) how such strong decay of flexural phonons could be observed? We will discuss both questions in the next sections.

\section{Absence of implication for the crumpling transition\label{Sec:StaticCor}}

The equilibrium stretching of membrane is determined by the condition that average displacement, cf. Eq. \eqref{eq:action:1},  vanishes in the absence of external tension 
\begin{equation}
    \langle \varepsilon_\alpha \rangle = \xi^2 - 1 + \sum_{\omega,\bm{k}} k_\alpha^2 \langle |h_{\bm{k},\omega}|^2\rangle = 0.
\end{equation}
This equation determines dependence of the stretching factor $\xi^2$ on temperature as
\begin{gather}
 \xi^2 = 1 - \frac{1}{2}\langle [\nabla h(\bm{x},t)]^2\rangle.
\end{gather}
The temperature $T_c$, at which $\xi^2$ vanishes, determines the crumpling transition of a membrane from the flat to crumpled phase. Computing $\langle h^2(\bm{x},t)\rangle$, we find 
\begin{gather}
 \langle [\nabla h(\bm{x},t)]^2\rangle = \int \frac{d\omega}{\pi}\int \frac{d^2\bm{k}}{(2\pi)^2} k^2 \Im \mathcal{G}_{\bm{k}}^R(\omega)  \coth \frac{\omega}{2T} \notag \\
 \simeq  2T \int \frac{d^2\bm{k}}{(2\pi)^2} k^2
  \int \frac{d\omega}{\pi} \frac{\Im \mathcal{G}_{\bm{k}}^R(\omega)}{\omega} 
  \notag \\
  = 2T \int \frac{d^2\bm{k}}{(2\pi)^2} k^2
\re \mathcal{G}_{\bm{k}}^R(0) 
=  \int \frac{d^2\bm{k}}{(2\pi)^2} \frac{2T k^2}{\varkappa k^{4-\eta}q_*^\eta} .
\label{eq:nablah2}
\end{gather}
It is exactly the same result as in the static theory. Therefore, the attenuation of flexural phonons does not affect the crumpling transition. Similarly, one can demonstrate that all the other static effects known as anomalous elasticity are not affected by the phonon dynamics.

\section{Time-dependent pair correlation function of out-of-plane displacement\label{Sec:Time}}

In this section we discuss the time-dependent pair correlation function of the out-of-plane displacement $h(\bm{x},t)$. We start from the variance,
$\langle h^2(\bm{x},t) \rangle$. 
As it follows from Eq. \eqref{eq:nablah2}, $\langle h^2(\bm{x},t) \rangle$ diverges in the infrared such that 
\begin{equation}
    \langle h^2(\bm{x},t) \rangle  \propto T L^{2\zeta} q_*^{-\eta}/\varkappa .
    \label{eq:hh:0}
\end{equation}
Here $L{\gg}1/q_*$ stands for the membrane's system size. The roughness exponent equals \cite{Doussal2018}
\begin{equation}
    \zeta= 1-\eta/2 . 
    \label{eq:z:dyn}
\end{equation}

Next we consider different time pair correlation function
\begin{align}
   \bigl \langle [h(\bm{x},t)-h(\bm{x},0)]^2\bigr \rangle  
   & = 2 \int \frac{d\omega}{\pi}\int \frac{d^2\bm{k}}{(2\pi)^2} 
   \sin^2\frac{\omega t}{2} \coth \frac{\omega}{2T}\notag \\
   & \times \Im \mathcal{G}_{\bm{k}}^R(\omega) . 
   \label{eq:HH:def}
\end{align}
Here the integrals are convergent both in ultra-violet and infra-red. So we consider infinite membrane. Then integral over $k$ is dominated by $k{\sim} [\rho \omega^2/(\varkappa q_*^\eta)]^{1/(4{-}\eta)}$ that corresponds to the mass-shell condition $\omega_k{=}\omega$. Therefore, we find 
\begin{align}
   \bigl \langle [h(\bm{x},t)-h(\bm{x},0)]^2\bigr \rangle  
   & \simeq 2 W_\eta \frac{T}{\rho} \left ( \frac{\rho}{\varkappa q_*^\eta}\right )^{\frac{2}{4-\eta}}\int_0^\infty \frac{d\omega}{\pi\omega}  \notag \\
   & \times  \omega^{-\frac{4-2\eta}{4-\eta}}
   \sin^2\frac{\omega t}{2}  ,
\end{align}
where the constant 
\begin{equation}
W_\eta {=} \!\int\limits_0^\infty \! \frac{dx\, x^{(3{-}\eta/2)} \mathcal{F}_2(x^{{-}2{+}\eta/2})/\pi}{[1{-}x^2(1{-}\mathcal{F}_1(x^{{-}2{+}\eta/2}))]^2{+}x^{4{-}\eta} \mathcal{F}_2^2(x^{{-}2{+}\eta/2}) } .
\end{equation}
Integral over frequency is dominated by $\omega{\sim}1/t$ such that we find 
\begin{equation}
\bigl \langle [h(\bm{x},t)-h(\bm{x},0)]^2\bigr \rangle 
\simeq  \widetilde{W}_\eta \frac{T}{\varkappa q_*^2} [\omega_{*} t]^{2\zeta/\textsf{z}} ,
\label{eq:hh:2}
\end{equation}
where $\widetilde{W}_\eta{=} {-}2^{2\zeta/z}\cos(\pi \zeta/z) \Gamma({-}2\zeta/z)W_\eta/\pi$. 
 We note the exact relation between dynamical and roughness exponents, 
\begin{equation}
    \textsf{z}=1+\zeta=2-\eta/2  .
    \label{eq:z:final}
\end{equation}
The result \eqref{eq:hh:2} is valid for long times $\omega_* t {\gg} 1$. Since the exponent $2\zeta/\textsf{z}{=}(2{-}\eta)/(2{-}\eta/2){<}1$, Eq. \eqref{eq:hh:2} implies a subdiffusive dynamics of out-of-plane deformations.  

One can combine Eqs. \eqref{eq:hh:0} and \eqref{eq:hh:2} in the following form, 
\begin{equation}
\bigl \langle [h(\bm{x},t)-h(\bm{x},0)]^2\bigr \rangle = \frac{T}{\varkappa} L^{2\zeta} q_*^{-\eta} \Xi\bigl( \omega_* t/(q_*L)^{\textsf{z}}\bigr ) ,
\label{eq:tt-d-hh}
\end{equation}
where the scaling function $\Xi(y)$ has the following asymptotic behavior,
\begin{equation}
\Xi(y) \propto \begin{cases}
{\rm const}, & \quad y\to 0 , \\
y^{2\zeta/\textsf{z}}, & \quad y\to \infty .
\end{cases}
\end{equation}

For shorter times, $T^{-1}{\ll}t {\ll} \omega_*^{-1}$, the integral over the momentum in Eq. \eqref{eq:HH:def} is still dominated by the mass shell condition. Since there is no renormalization of the bending rigidity for $k{\gg} q_*$, we find
diffusive-type dynamics at $T^{-1}{\ll}t {\ll} \omega_*^{-1}$,
\begin{equation}
\bigl \langle [h(\bm{x},t)-h(\bm{x},0)]^2\bigr \rangle 
\sim  \frac{T}{\varkappa q_*^2} \omega_{*} t .
\label{eq:hh:22}
\end{equation}

We discuss significance of the above results in the next Section.  

\section{Discussion and conclusion \label{Sec:DiscConc}}

\subsection{Comparison with the generalized Langevin approach}

 One could try to describe the low frequency dynamics of the 2D membrane phenomenologically by means of the Langevin-type approach. The form \eqref{eq:Green:main} of the exact Green's function for the out-of-plane displacement $h$ at low frequencies suggests the following Langevin-type equation
\begin{equation}
{-} \rho \tilde{D} k^2 (q_*/k)^{\eta/2} \partial_t h_{\bm{k}}(t) {=}  \rho \omega_{\bm{k}}^2 h_{\bm{k}}(t)  {+} k (q_*/k)^{{\eta}/{4}} f_{\bm{k}}(t) .
\label{eq:Langevin}
\end{equation}
Here $\tilde{D}{=}\mathcal{F}_2(0) D$ and
a white-noise random force has the correlation function dictated by the fluctuation-dissipation relation,
\begin{equation}
\langle f(\bm{x},t) f(\bm{x}^\prime,t^\prime) \rangle = 4 T \tilde{D} \rho  \delta(t-t^\prime)\delta(\bm{x}-\bm{x}^\prime) .
\end{equation}
We note that in contrast with Langevin-type equation used in Refs. \cite{Mizuochi2014,Granato2022,Granato2023,Steinbock2023}, all terms of Eq. \eqref{eq:Langevin} contains explicit $k$-dependence. We emphasize that Eq. \eqref{eq:Langevin} can be only used for study of long time dynamics, $\omega_{\bm{k}} t{\gg} 1$, where $\omega_{\bm{k}}$ is fixed by the magnitude of a relevant wave vector, $k{\sim}1/L$. In general, one could try to derive the Langevin-type equation for the considered problem with the help of Wyld technique (see Ref. \cite{Lebedev2012} for details) or, alternatively, by means of the Keldysh path integral. We leave it for future works.

Another complication with application of Langevin-type equation to description of dynamics of a 2D membrane is nonlinearity (interaction of flexural phonos) which leads not only to renormalization of the bending rigidity and attenuation but also to real mode coupling \cite{Croy2012,De2020}. The latter appears as nonlinear terms in the Langevin-type equation.

\subsection{Attenuation of flexural phonons for membranes of higher dimensions}

The Wyld technique has recently been used for analysis of classical dynamics (in the sense of inequality $\omega{\ll}T$) of  a $D{=}4{-}\epsilon$ dimensional crystalline membrane \cite{Lebedev2012}. Analysing the perturbative renormalization group controlled by a small parameter $\epsilon{\ll}1$, the authors of Ref. \cite{Lebedev2012} led essentially to the same scaling form of the Green's function, cf. Eq. \eqref{eq:Green:main}, and to the same expression for the dynamical exponent $\textsf{z}$, cf. Eq. \eqref{eq:z:dyn}. Together with our result, this suggests that the scaling form \eqref{eq:Green:main} and Eq. \eqref{eq:z:dyn} for $\textsf{z}$ are valid for a membrane of arbitrary dimension $D{\geqslant}2$.

\subsection{Attenuation of in-plane phonons}

Due to the $O(2)$ rotational symmetry existing for a membrane in the flat phase \cite{Aronovitz1989}, the renormalization of in-plane phonons is intimately related with that of flexural phonons, cf. Eq. \eqref{eq:ren:12}. In order to find the spectrum of in-plane phonons at low momentum, we use the relation $\omega{\sim} k (Y/\rho)^{1/2}$ in which $Y$ is substituted by $1/\Pi^R_k(\omega)$. We note that we do not distinguish between longitudinal and transverse in-plane phonons. Since, as we will check below, the frequency of in-plane phonons is parametrically higher than that of flexural phonons, one needs to employ large frequency asymptote of the polarization operator, Eq. \eqref{eq:PolOp:main:A2}. Then we find that the spectrum of the longitudinal and transverse in-plane phonons (at $k{\ll}q_*\sqrt{T/\varkappa}$) is given as 
\begin{equation}
    \omega_k^{(l,t)}{=} s_{l,t} \omega_* \left (\frac{k \sqrt{\varkappa}}{q_* \sqrt{T}}\right )^{\textsf{z}^\prime} ,\quad 
    \textsf{z}^\prime=\frac{2{-}\eta}{1{+}\eta/2} .
    \label{eq:inplane:freq}
\end{equation}
 Here $s_{l,t}$ are  some complex numbers.
We note that the region of validity of Eq. \eqref{eq:inplane:freq} is determined by the inequality $Y \Pi^R_k(\omega_k^{(l,t)}){\gg} 1$. Also we note that the assumption $\omega_k^{(l,t)}{\gg}\omega_k$ is satisfied indeed.  At $k{\gg}q_*\sqrt{T/\varkappa}$ the spectrum of the in-plane phonons is not renormalized, $\omega_k^{(l,t)}{\sim} k$.
Since at such momenta, $\omega_{k}^{(l,t)} {\sim} k \sqrt{Y/\rho} {\gg} \omega_*$, 
we use the following estimate  
in this region: $Y \im \Pi^R_k(\omega_k^{(l,t)}) {\sim}  q_* \sqrt{T}/(k\sqrt{\varkappa})  {\ll} 1 $. Thus, attenuation becomes
\begin{equation}
    \frac{1}{\tau_{\bm{k}}}\simeq\frac{\rho (\omega_{\bm{k}}^{(l,t)})^2}{\rho \omega_{\bm{k}}^{(l,t)} } \left(\frac{q_{*} \sqrt{T}}{k\sqrt{\varkappa}}\right) \simeq \omega_*.
\end{equation}

We emphasize that in contrast with the case of flexural phonons, the scaling of frequency with momentum in Eq. \eqref{eq:inplane:freq} is different from that one could envision on the basis of static renormalization of elastic moduli \eqref{eq:ren:12}. 

We expect validity of the result \eqref{eq:inplane:freq} for the dynamical exponent of the in-plane phonons for membranes of an arbitrary dimension $D{\geqslant}2$. For $D{=}4{-}\epsilon$ our prediction contradicts to the result of Ref.~\cite{Lebedev2012}. We believe that the origin for such a discrepancy is that the static renormalization of elastic moduli \eqref{eq:ren:12} was used in Ref.~\cite{Lebedev2012} to derive the spectrum of in-plane phonons.

\subsection{Flexural phonon attenuation beyond the universal regime}

In the above discussion we consider the universal region of small frequency and momenta, $k{\ll}q_*$ and $\omega{\ll}\omega_*$. Although, such a situation realizes typically in experiments, it is worthwhile to discuss the behavior of the imaginary part of the self energy at large wave vectors, $q_*{\ll}k{\ll}q_T{=}q_*/\sqrt{g}$ and frequencies, $\omega_*{\ll}|\omega|{\ll}T$ (regions II and III in Fig. \ref{fig:UniversalRegime}). The estimates given in Appendix \ref{App:I} result in the following behavior 
\begin{equation}
\im \Sigma_{\bm{k}}^R(\omega) {\sim} 
\frac{T^2 Y^2}{\varkappa^3}
\setstretch{2.0}
\begin{cases}
\displaystyle \frac{D k^{2{+}\eta}}{\omega q_*^\eta} , \,\, \textrm{II$_a$}:\, \omega {\gg} \omega_* {\gg} D k^2 , \\
\displaystyle \frac{Dk^2}{\omega} , \,\, \textrm{III$_a$:}\, \omega {\gg} D k^2 {\gg} \omega_* , \\
\displaystyle 
\frac{\omega}{Dk^2}, \,\, \textrm{II$_b$ \& III$_b$:}\, \omega {\ll} D k^2 , \, q_*{\ll} k .
\end{cases}
\label{eq:nonUn:1}
\end{equation}
Interestingly, near the mass shell, the imaginary part of the self energy is enhanced  by a factor $k/q_*{\gg} 1$  (see Fig. \ref{fig:UniversalRegime}), 
\begin{equation}
   \im \Sigma_{\bm{k}}^R(\omega) \sim \frac{T^2 Y^2}{\varkappa^3} \frac{k}{q_*} , \quad |\omega-Dk^2|\ll \omega_* .
   \label{eq:nonUn:2}
\end{equation}
Estimating the 
 attenuation coefficient 
of the flexural phonon with the momentum $q_*{\ll}k{\ll}q_T$ as
$1/\tau_k{=} \im \Sigma_{\bm{k}}^R(\omega_k^{(0)})/(\rho \omega_k^{(0)})$, we find that $\omega_k^{(0)}\tau_k{\sim}(k/q_*)^3 {\gg} 1$. Therefore, 
there is almost no attenuation of the spectrum of flexural phonons with high momenta $q_*{\ll}k{\ll}q_T$.

\begin{figure}[t]   
\centerline{\includegraphics[width=\columnwidth]{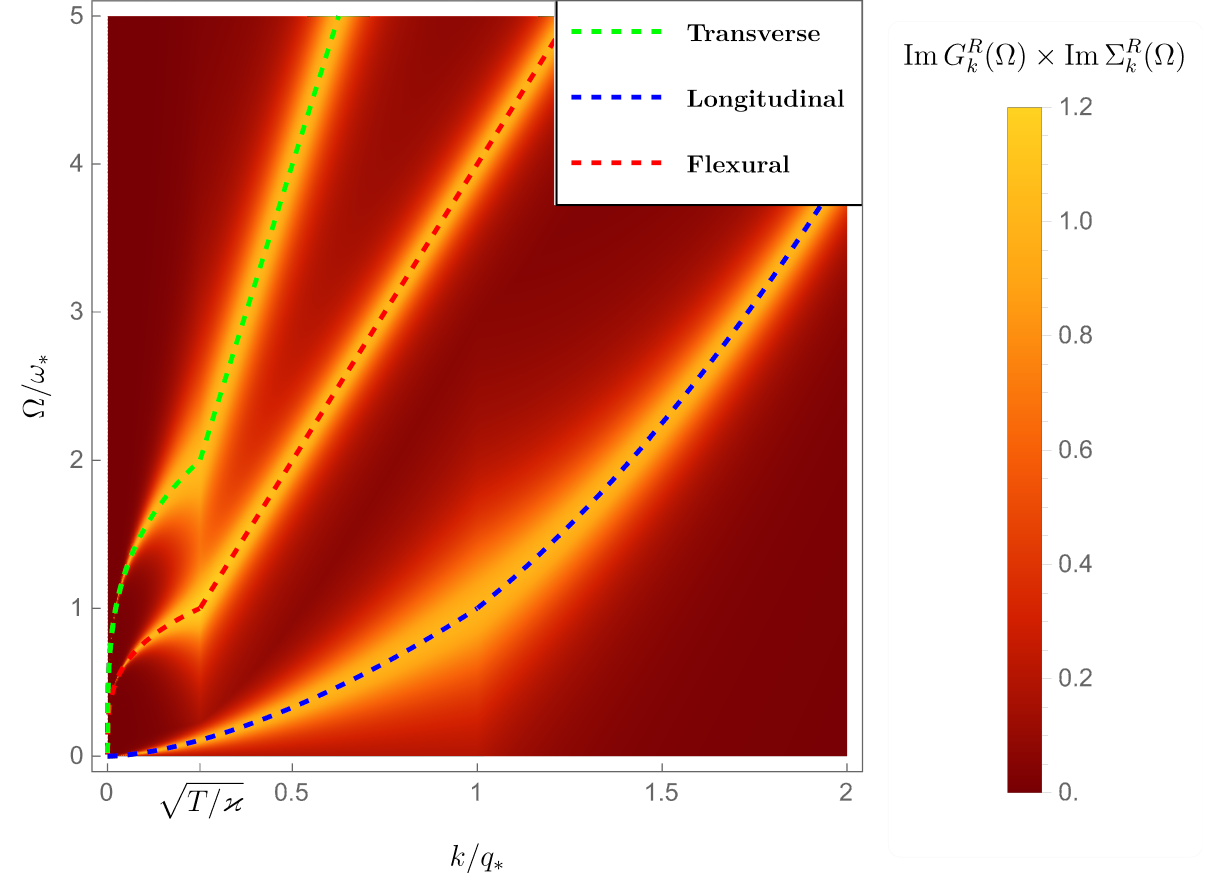}}
    \caption{Sketch of the dependence of the phonon spectral function on the momentum and frequency. For convenient normalization, we plot $\Im \mathcal{G}^R_{\bm{k}}(\omega) \Im \Sigma^R_{\bm{k}}(\omega)$. We emphasize that change of the spectrum of flexural phonons occurs at $k{\sim}q_*$ while for the in-plane phonons it happens at $k{\sim}q_*\sqrt{T/\varkappa}$.}
    \label{pic:Spectra}
\end{figure}

\subsection{Benchmarking against experiments}

In this paper we present detailed microscopic theory of phonon attenuation in two-dimensional flexible materials. We mention that the phonon spectrum in graphene has recently been  measured by the method of the high resolution electron energy loss spectroscopy \cite{Li2023}. We note that experimental data demonstrate some broadening of phonon spectrum. However, in order to perform detailed benchmarking of our theory, more detailed experimental data of the spectrum around the $\Gamma$ point  is needed.

\subsection{Implications for mechanical nanoresonators}

It is instructive to estimate numerical magnitudes of important parameters in our theory. Having in mind graphene as an example of two-dimensional crystalline membrane, we find that the Ginzburg length is $1/q_*{\approx} 1$~nm and $\omega_*{\approx} 200$~GHz. For $L{\gg}1/q_*$, the frequency of typical out-of-plane deformation can be estimated as $\omega_{k{\sim}1/L}{\approx} 3$ MHz for a typical size $L{=}1$ $\mu$m. Also, Eq. \eqref{eq:hh:0} allows us to estimate typical amplitude of the flexural deformations as $h{\sim}10$ nm for the same $L{=}1$~$\mu$m.

Recently, the measurement of time-dependent out-of-plane fluctuations in graphene has been performed by means of scanning tunneling spectroscopy \cite{Ackerman2016}. In agreement with our theory, the long-time dynamics characterized by the pair correlation function $\langle [h(\bm{x},t){-}h(\bm{x},0)]^2\rangle$ was found to be subdiffusive. However, the corresponding exponent was estimated to be equal $0.3$ in contrast to our prediction $2\zeta/\textsf{z}{\simeq}0.75$. Several possible reasons for such a discrepancy might be proposed. At first, the data in the experiment of Ref. \cite{Ackerman2016} contains two types of fluctuations: fast small amplitude fluctuations and slow large amplitude excursions. While the former can be assumed to be the thermal fluctuations studied in our paper, the latter were related with spontaneous changing of local curvature. Such local buckling is not taken into account in our theory. Secondly, the scanning tunneling microscopy tip can induce a local tension that affects the dynamics of thermal fluctuations. 
At third, the experimental data presented in Ref. \cite{Ackerman2016} were collected from multiple graphene membranes. It is known \cite{Gornyi2016} that a quenched random curvature is important for graphene samples. Different graphene flakes in the experiment of Ref. \cite{Ackerman2016} could have a different realization of a quenched random curvature (due to some disorder).
Therefore, one needs to study the dynamics of flexural phonons in the presence of disorder. 
We leave more detailed investigation of the effects discussed above for future work.

\subsection{The effect of a non-zero tension}

The theory presented in this work was developed for free-standing materials  in the absence of the tension,  $\sigma{\equiv}0$. However, if membrane is lying on a substrate with a hole then the substrate imposes a stress on a part of the membrane above the hole. Therefore, the membrane experiences a nonzero tension $\sigma$.

As well-known \cite{Aronovitz1988,Aronovitz1989,Radzihovsky1992}, a small tension, $\sigma{\ll} \sigma_*{=}\varkappa q_*^2$, (i) suppresses the renormalization of the bending rigidity at wave vectors $k{<}q_\sigma$, where $q_\sigma{=}q_*(\sigma/\sigma_*)^{1/(2{-}\eta)}$ and (ii) transforms the spectrum of flexural phonons into sound-like one, $\omega_k^{(\sigma)}{=} k \sqrt{\sigma/\rho}$. Therefore, there is no surprise that tension affects the attenuation of flexural phonons. In particular, one can derive the following estimate \cite{Kokovin2023s}
\begin{equation}
\im \Sigma_{\bm{k}}^R(\omega_k^{(\sigma)})\sim \varkappa_\sigma k^4 , \qquad k\ll q_\sigma , 
\label{eq:ImS:tension}
\end{equation}
where $\varkappa_\sigma{=}\varkappa (q_*/q_\sigma)^\eta$. 
Therefore, the decay 
rate of the flexural phonon at $k{\ll}q_\sigma$ becomes 
\begin{equation}
1/\tau_k^{(\sigma)}=\im \Sigma_{\bm{k}}^R(\omega_k^{(\sigma)})/(\rho\omega_k^{(\sigma)})\sim(k/q_\sigma)^2 \omega_k^{(\sigma)} \ll \omega_k^{(\sigma)} .
\end{equation}
Therefore, a nonzero tension results in parametric narrowing of the spectral line for the flexural phonon. Interestingly, the width of spectral line becomes temperature dependent  in the presence of a non-zero tension , $1/\tau_k^{(\sigma)}{\sim}T^{\alpha}$, where $\alpha{=}\eta/(2{-}\eta){\simeq}0.67$ \cite{Kokovin2023s}.

Finally, we note that there are other mechanisms for decay of the out-of-plane displacement dynamics in nanoelectromechanical resonators 
\cite{Bachtold2022}. Their discussion is beyond the scope of the present work.

\subsection{Summary} 

To summarize we studied the attenuation of the 
phonons in free-standing 2D crystalline membranes. We explored high temperature regime (relevant for experiments) in which  
flexural phonons can be treated classically, $T{\gg}\omega_k$. We found that in the universal regime, $k{\ll}q_*$, the broadening  of the flexural phonon spectral line  is of the order of the spectrum itself while at $q_*{\ll}k{\ll}q_T$ the broadening is  parametrically  suppressed. Focusing on the universal regime, we established the exact expression for the dynamical exponents $\textsf{z}$, see Eq. \eqref{eq:z:final},  and $\textsf{z}^\prime$, see Eq. \eqref{eq:inplane:freq}, for flexural and in-plane phonons, respectively.  We applied our theory to computation of the time-dependent pair correlation function of membrane's height and found its subdiffusive behavior at long times in qualitative accordance with the experiments. 
Finally, we discussed some future research directions.

\begin{acknowledgements}

The authors thank V. Kachorovskii for continuous interest to this work and for useful comments.  The authors are grateful to E. Kats and V. Lebedev for fruitful discussions.   The work was funded in part by the Russian Ministry of Science and Higher Educations and the Basic Research Program of HSE. 

\end{acknowledgements}


%
%
%
%
%
\appendix

\section{The effect of dynamical part of interaction between flexural phonons mediated by in-plane ones\label{Appendix:1:Inplane}}

In this Appendix we present an estimate for contribution of dynamical part of bare interaction between flexural phonons to $\Im \Sigma^R_{\bm{k}}(\omega)$.  

The self energy correction in the first order of perturbation theory is given by (see Ref. \cite{Burmistrov2016}. Eq. (B6)):
\begin{gather}
    \im \Sigma^{(m), R}_{\bm{k}}(\omega) {=}  T \omega \int\frac{d\Omega}{2\pi}\int\frac{d^2\bm{q}}{(2\pi)^2}\frac{\left((\bm{k}\cdot\bm{q}-q^2\right)^2}{q^4}(\bm{q}\cdot\bm{k})^2 \notag \\
    \times\frac{\im R^{(mmmm),R}_{\bm{q}}(\Omega)}{\Omega}\frac{\im G^{R}_{\bm{k} + \bm{q}}(\Omega + \omega)}{\omega + \Omega}
    \label{eq::A1}
\end{gather}

where 

\begin{equation}
R^{(mmmm)}_{\bm{q}}(i\Omega) = \rho \frac{(2\mu + \lambda)\Omega^2}{(2\mu + \lambda)q^2 + \rho \Omega^2}  .   
\label{eq::A2}
\end{equation}
This is one of the four additional interaction terms, but all of them have the same scaling properties. For conciseness, we will only evaluate this term.

As in the main text, we focus on  the region $k {\ll} q_{*}$. In this domain, $\omega_{\bm{k}} {=} D k^2 {\ll} c_l k {=} \varepsilon_{\bm{k}}$, where $c_l {=} \sqrt{(2\mu {+} \lambda)/\rho}$ denotes the speed of longitudinal sound mode.
We are mostly interested in $\omega {\sim} \omega_{\bm{k}}$, 
since the decay rate is determined by the frequency on the mass-shell. First, we find the imaginary part of the retarded interaction:
\begin{gather}
    \im R^{(mmmm),R}_{\bm{q}}(\Omega) = \frac{\pi (2\mu + \lambda)\Omega}{2}\sum_{s=\pm} \delta(\Omega + s \varepsilon_{\bm{q}}) .
    \label{eq::A3}
\end{gather}
The imaginary part of the retarded Green's function becomes
\begin{equation}
    \im G^{R}_{ \bm{q}}(\omega) = \frac{\pi }{2\rho \omega}\sum_{s=\pm} \delta(\omega + s \omega_{\bm{q}}) .
    \label{eq::A4}
\end{equation}

We then substitute Eqs. \eqref{eq::A3} and \eqref{eq::A4} into Eq. \eqref{eq::A1} and integrate over $\Omega$,  
\begin{gather}
    \im \Sigma^{(m), R}_{\bm{k}}(\omega) {=}\frac{c^2_l T \omega}{32\pi}\int d^2\bm{q}\frac{\left((\bm{k}\cdot\bm{q}-q^2\right)^2}{q^4} \frac{(\bm{q}\cdot\bm{k})^2}{\omega_{\bm{k}+\bm{q}}^2} \notag \\
    \times\sum_{s{=}{\pm}}\left[\delta(s\varepsilon_{\bm{q}} + \omega + \omega_{\bm{k} + \bm{q}}) + \delta(s\varepsilon_{\bm{q}} + \omega - \omega_{\bm{k} + \bm{q}})\right].
\end{gather}

We proceed by introducing new variables $\bm{k} {=} k \bm{n}$, $\bm{q} {=} k \bm{r}$, $z {=} \omega/\omega_{\bm{k}}$ and by making the expressions dimensionless. We also introduce the parameter $\alpha_k{=} \varepsilon_{\bm{k}}/\omega_{\bm{k}} {\gg} 1$. In terms of those variables, we obtain
\begin{gather}
   \im \Sigma^{(m), R}_{\bm{k}}(\omega) {=}\frac{T}{\varkappa}\frac{\rho \omega_{\bm{k}} \omega}{32\pi} \int d^2\bm{r}\frac{\left((\bm{r}\cdot\bm{n})-r^2\right)^2}{r^4} \frac{\alpha_k^2 (\bm{r}\cdot\bm{n})^2}{|\bm{r}+\bm{n}|^4} \notag \\
   \times\sum_{s{=}{\pm}1}\left[\delta(s\alpha_k r {+} z {+} |\bm{r} {+} \bm{n}|^2) + \delta(s\alpha_k r {+} z {-} |\bm{r} {+} \bm{n}|^2)\right]. 
\end{gather}

At large $\alpha_k$ and $z{\sim}1$, the argument of $\delta$-function is zero either at $r{\sim}\alpha_k{\gg}1$ or $r{\sim}1/\alpha_k{\ll}1$. In both cases, we can approximate $|\bm{r} {+} \bm{n}|^2{\sim}r^2{+}1$. 

Then the integral over the angle can be evaluated separately: 
\begin{equation}
   \int\limits_0^{2\pi} d\varphi \frac{\left(\bm{r}\cdot\bm{n}{-}r^2\right)^2}{r^4} \frac{(\bm{r}\cdot\bm{n})^2}{|\bm{r}{+}\bm{n}|^4} \simeq 
   \begin{cases}
 \frac{3\pi}{4}, &  r{\ll} 1 ,
 \\
 \frac{\pi}{r^2}, &  r{\gg} 1 .
 \end{cases}
 \label{eq::A7}
\end{equation}

We then integrate over $r$ using asymptotics \eqref{eq::A7} and obtain for $z>0$:
\begin{gather}
     \im \Sigma^{(m), R}_{\bm{k}}(\omega) {=} \frac{T}{32\varkappa}\rho \omega_{\bm{k}} \omega\Bigl [2 + \frac{3}{4}(|z-1|+z+1)\Bigr ].
\end{gather}

We can see, that the correction is small in virtue of the small parameter $T/\varkappa{\ll} 1$. 
The same parameter controls other corrections occurring from dynamics of the in-plane phonons.

\section{Evaluation of the static limit of the diagram in Fig. \ref{pic:SelfEnergy}\label{Appendix:2:Non-SCSA-diag:static}}

In this Appendix we demonstrate how the static limit of the diagram shown in Fig. \ref{pic:SelfEnergy} transforms into Eq. \eqref{eq:diag:Fig3:static} at high temperatures. The aforementioned diagram is given by
\begin{gather}
\Sigma_{\bm{k}}^{(2)}(0){=}{-}4 \sum_{\bm{q}, \bm{Q}, \Omega, \omega}  S(\bm{q}, \bm{Q}) G_{\bm{k}{+}\bm{q}}(i\omega) G_{\bm{k}{+}\bm{Q}}(i\Omega) \notag \\
{\times} G_{\bm{k}{+}\bm{q}{+}\bm{Q}}(i\omega + i\Omega) N_{\bm{q}}(i\omega)N_{\bm{Q}}(i\Omega) ,
\label{eq::B1}
\end{gather}
where for convenience we introduced
\begin{equation}
    S(\bm{q}, \bm{Q}) = \frac{[(\bm{k}{+}\bm{Q}) {\times} \bm{q}]^2}{q^2}  \frac{[(\bm{k}{+}\bm{q}) {\times} \bm{Q}]^2}{Q^2} 
\frac{[\bm{k} {\times} \bm{q}]^2}{q^2} \frac{[\bm{k} {\times} \bm{Q}]^2}{Q^2} .
\end{equation}
At first, we transform the sum over bosonic Matsubara frequencies $\omega$ into the integral along the real axis
\begin{gather}
\Sigma_{\bm{k}}^{(2)}(0){=}{-}4  \sum_{\bm{q}, \bm{Q}, \Omega}  S(\bm{q}, \bm{Q})N_{\bm{Q}}(i\Omega)  G_{\bm{k}{+}\bm{Q}}(i\Omega) \notag 
\int \frac{d\omega}{2\pi}
\\
\times \coth \left(\frac{\omega}{2T}\right) \Bigl[ G_{\bm{k}{+}\bm{q}{+}\bm{Q}}(\omega {+} i\Omega) \im \bigl(N^R_{\bm{q}}(\omega)G^R_{\bm{k}{+}\bm{q}}(\omega)\bigr)
\notag \\
+ N_{\bm{q}}(\omega-i\Omega)G_{\bm{k}{+}\bm{q}}(\omega-i\Omega) \im G^R_{\bm{k}{+}\bm{q}{+}\bm{Q}}(\omega)
\Bigr ].
\label{eq::B3}
\end{gather}
Next, similarly, we transform the sum over bosonic Matsubara frequencies $\Omega$ into the integral over real axis,
\begin{gather}
\Sigma_{\bm{k}}^{(2)}(0){=}{-}4  \sum_{\bm{q}, \bm{Q}}  S(\bm{q}, \bm{Q})\int \frac{d\Omega d\omega}{(2\pi)^2} \coth\left(\frac{\Omega}{2T}\right) \coth\left(\frac{\omega}{2T}\right)
\notag \\
\times\Bigl [ \im \bigl (N^R_{\bm{Q}}(\Omega)  G^R_{\bm{k}{+}\bm{Q}}(\Omega)
G^R_{\bm{k}{+}\bm{q}{+}\bm{Q}}(\omega {+} \Omega)\bigr ) \notag \\
\times
 \im \bigl (N^R_{\bm{q}}(\omega)G^R_{\bm{k}{+}\bm{q}}(\omega)\bigr ) 
 + \im G^R_{\bm{k}{+}\bm{q}{+}\bm{Q}}(\omega) \notag \\
 \times 
 \im \bigl (N^R_{\bm{Q}}(\Omega)  G^R_{\bm{k}{+}\bm{Q}}(\Omega)N^A_{\bm{q}}(\omega-\Omega)G^A_{\bm{k}{+}\bm{q}}(\omega-\Omega)
 \bigr )
 \Bigr ].
\label{eq::B3}
\end{gather}
In the high-temperature regime ($T{\gg} |\Omega|, |\omega|$) the hyperbolic cotangent  can be replaced by the first term in it's Taylor series. Then using Kramers--Kronig relations, we perform integrals over $\omega$ and $\Omega$, and, thus, we derive \eqref{eq:diag:Fig3:static}.

\section{Formulation of dynamical perturbation theory \label{Appendix:3:Dyn:Pert}}

In this Appendix we demonstrate how the perturbation theory around the Green's function \eqref{eq:GreenG:bare:2} can be formulated in a regular fashion. 

The ``bare" Green's function is related to the Green's function defined in Eq. \eqref{eq:GreenG:bare:2} by the Dyson equation with static self energy,
\begin{equation}
    G^{-1}_{\bm{k}}(i\omega) = \left[\mathcal{G}^{-1}_{\bm{k}}(i\omega) +\Sigma_{\bm{k}}(0)\right]^{-1} 
    \label{eq::C1}
\end{equation}

We can also rewrite this equation in terms of the infinite series
\begin{equation}
    G_{\bm{k}}(i\omega) = \mathcal{G}_{\bm{k}}(i\omega) - \mathcal{G}_{\bm{k}}(i\omega)\Sigma_{\bm{k}}(0)\mathcal{G}_{\bm{k}} (i\omega) + \dots
    \label{eq::C2}
\end{equation}

\begin{figure}
    \centering
    \includegraphics[width =0.85\columnwidth]{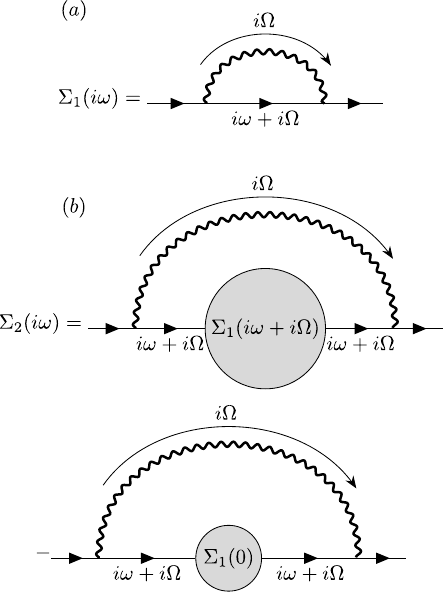}
    \caption{(Top) First-order of interaction self energy correction, (Middle) Second-order main term and (Bottom) Second-order extra term. Wiggly lines denote screened interaction \eqref{eq:N:screened}, solid lines denote Green's functions \eqref{eq:GreenG:bare:2}}.
    \label{fig:Appendix:C}
\end{figure}

To compute the dynamical self energy corrections, we need to insert the series from Eq. \eqref{eq::C2} into the series for $\Sigma_{\bm{k}}(i\omega) - \Sigma_{\bm{k}}(0)$ (See Fig, \ref{fig:Appendix:C}). Calculation of the imaginary part, $\im (\Sigma^R_{\bm{k}}(i\omega) {-} \Sigma^R_{\bm{k}}(0))$, is more convenient because $\im \Sigma^R_{\bm{k}}(0) {=} 0$.

The first term in the right hand side of Eq.  \eqref{eq::C2} forms a sequence of self energy diagrams, where all the ``bare" Green's functions are substituted by the one from Eq. \eqref{eq:GreenG:bare:2}. These contributions are termed as the main terms. The next terms in the right hand side of Eq. \eqref{eq::C2} produce additional set of diagrams (extra terms).  

The above statements can be illustrated by diagrams in Fig. \ref{fig:Appendix:C}. In Fig. \ref{fig:Appendix:C}a, the diagram of the first order in dynamically screened interaction is shown. There a solid line in the ``bare'' Green's function $G$. Substituting it by the second term in the right hand side of  \eqref{eq::C2}, we produce the second diagram in Fig. \ref{fig:Appendix:C}b. We note that the extra diagram is formally of infinite order in 
dynamically screened interaction, since it involves  the exact static self energy. In the same way diagrams of the second order in the dynamically screened interaction produces extra diagrams with the exact static self energy.

Now we argue that extra terms does not spoil the scaling of the main self energy corrections in the universal regime $q {\ll} q_{*}$. Indeed, the static self energy, $\Sigma_{\bm{k}} (0) {=} \varkappa k^{4{-}\eta} q_{*}^{\eta}$, has the same $k$-dependence as the dynamical self energy at the mass shell, $\omega{=}\omega_k$. Similarly to the main terms, the frequency integrals in extra terms are still dominated by the frequencies corresponding to the mass shell conditions.   
Therefore, extra terms produce the same scaling between frequency and momentum as the main ones.

The same argument allows us to work with screened Green's functions \eqref{eq:GreenG:bare:2} (instead of the ``bare'' ones) in the polarization operator \eqref{eq:pol:R:im}.

\section{Computation of asymptotic expression for 
the polarization operator \eqref{eq:P000}
\label{Appendix:4:PolOper:0}
}

In this Appendix we present derivation of Eqs. \eqref{eq:ImP:small:as} and \eqref{eq:ImP:high:as}.
Using Eq. \eqref{eq::A4} for the imaginary part of the retarded Green's function, we obtain
from Eq. 
\eqref{eq:pol:R:im},
\begin{gather}
\im \Pi^R_{\bm{q}}(\Omega) = 
\frac{\pi T\Omega}{12 q^{2\eta}_*} \int \frac{d^2\bm{k}}{(2\pi)^2}
\frac{[\bm{k}\times \bm{q}]^4}{q^4}
\frac{1}{\varkappa k^{4-\eta}} \frac{1}{\varkappa|\bm{k}+\bm{q}|^{4-\eta}}
\notag \\
\times \sum_{s=\pm 1} 
\Bigl [ \delta \bigl (s\Omega {+}\omega_k{-}\omega_{\bm{k}{+}\bm{q}}\bigr )
+ \delta \bigl (s\Omega {+}\omega_k{+}\omega_{\bm{k}{+}\bm{q}}\bigr )
\Bigr ] .
\end{gather}

In order to calculate this integral, it is convenient to introduce new variable $\bm{y} {=} |\bm{k} {+} \bm{q}|$. We note, that the Jacobian of this transformation is
\begin{equation}
    \left|\frac{\partial(k, \varphi)}{\partial (k,y)}\right| = \frac{2y}{k q \sin \varphi} .
\end{equation}
The factor of 2 emerges because the integrand is an even function of the
angle, and our substitution is single-valued only in one half-plane. In terms of new variables integral takes the form
\begin{gather}
    \operatorname{Im} \Pi^R_{\bm{q}}(\Omega) =\frac{T}{24 \pi} \frac{\Omega }{\varkappa^2 q_*^{2\eta} q} \int {y d y d k} \frac{k^{\eta} \sin ^3 \varphi }{y^{4-\eta}} \notag 
    \\
    \times\sum_{s=\pm}\left[\delta\left(s\Omega+\omega_{\bm{k}}-\omega_{\bm{y}}\right)+\delta\left(s\Omega+\omega_{\bm{k}} +\omega_{\bm{y}}\right)\right] .
\end{gather}
We then proceed by making the integral dimensionless by introducing  variables $z {=} \Omega/\omega_{\bm{q}},\;  x {=} k/q, \; a {=} y/q$. We also rewrite $\sin^3 \varphi$ in terms of new variables. Then we obtain
\begin{gather}
     \operatorname{Im} \Pi^R_{\bm{q}}(\Omega) {=}\frac{T}{24 \pi} \frac{z }{\varkappa^2 q_*^{2\eta} q^{2-2\eta}}\int_0^{\infty}dx \int_{|x-1|}^{|x+1|} da \frac{x^{\eta}}{a^{3-\eta}}\sum_{s=\pm}\notag \\
 {\times}\left[\delta\left(s z{+}x^{2{-}\eta/2}{-}a^{2{-}\eta/2}\right){+}\delta\left(s z{+}x^{2{-}\eta/2}{+}a^{2{-}\eta/2}\right)\right]  \notag 
 \\
 \times \left(1 - \left(\frac{x}{2}+\frac{1}{2x}-\frac{a^2}{2x}\right)^2\right)^{3/2} . 
\end{gather}
The integral in this form can be evaluated in different limits. For $z \to 0$, we find
\begin{gather}
 \operatorname{Im} \Pi^R_{\bm{q}}(\Omega) {=}\frac{T}{24 \pi} \frac{z }{\varkappa^2 q_*^{2\eta} q^{2-2\eta}}\int_0^{\infty}dx \int_{|x-1|}^{|x+1|} \frac{x^{\eta}}{a^{3-\eta}}\notag \\
 \times\left(1 {-} \left(\frac{x}{2}{+}\frac{1}{2x}{-}\frac{a^2}{2x}\right)^2\right)^{3/2}\left[\delta\left(x^{2-\eta/2}-a^{2-\eta/2}\right)\right] .  
\end{gather}
This integral can be computed exactly. This way we obtain ($\mathcal{P}^{(0)}_2(z)$ is defined according to Eq. \eqref{eq:P000})
\begin{gather}
\mathcal{P}^{(0)}_2(z) \simeq C_\eta^{(0)} z, \notag \\
C_\eta^{(0)} =\frac{2^{2{-}3\eta/2} \Gamma^2(2{-}\eta/2)\Gamma(3/2{+}\eta/2)\Gamma(3/2{-}5\eta/4)}{(1{-}\eta/4)\Gamma(1{+}\eta/2)\Gamma(1{-}\eta)\Gamma(4{-}5\eta/4)} 
\end{gather}
In the opposite limit $z{\gg}1$ the integral is somewhat more complicated. In this case we will return to the representation of the integral in terms of the momentum and the angle

\begin{gather}
\operatorname{Im} \Pi^R_{\bm{q}}(\Omega) {=}\frac{T}{24 \pi} \frac{z }{\varkappa^2 q_*^{2\eta} q^{2-2\eta}}\int_0^{\infty}dx x^{\eta + 1}  \int_{0}^{\pi} d\varphi
 \sin^4{\varphi}
\notag \\
\times\frac{\delta\left(z-x^{2-\eta/2}-(x^2 + 1 + 2x \cos{\varphi})^{1-\eta/4}\right)}{(x^2 + 1 + 2x \cos{\varphi})^{2-\eta/2}} .
\end{gather}
The other delta functions were excluded because they give subdominant contributions to asymptotics.
It can be seen that due to $\delta$-function, integration over $x$ sets $x^{2{-}\eta/2} 
{\approx} z/2$. Hence, we find 
\begin{gather}
    \operatorname{Im} \Pi^R_{\bm{q}}(\Omega) {=}
    \frac{T}{24 \pi (4{-}\eta)} \frac{z}{\varkappa^2 q_*^{2\eta} q^{2-2\eta}}\int\limits_{0}^{\pi} d\varphi\frac{(z/2)^{\frac{3\eta}{4{-}\eta}}}{(z/2)^{2}}\sin^4{\varphi}
\end{gather}
Integrating over $\varphi$, we obtain
\begin{gather}
\mathcal{P}^{(0)}_2(z) \simeq C_\eta^{(\infty)} z^{-(1-\eta)/(1-\eta/4)},\notag \\
C_\eta^{(\infty)} =\frac{2^{\eta+(1-\eta)/(1-\eta/4)} \sqrt{\pi}\Gamma^2(2-\eta/2)\Gamma(3/2+\eta/2)}{(4-\eta)\Gamma(1+\eta/2)\Gamma(1-\eta)}  
\label{eq:P2:zL:new}
\end{gather}
In order  to find the real part of the polarization operator, $\mathcal{P}^{(0)}_1(z)$, we will use the Kramers-Kronig relation
\begin{gather}
\mathcal{P}^{(0)}_1(z) =1+\frac{2 z^2}{\pi} {\rm p.v.} \int\limits_0^\infty \frac{d y}{y} \frac{\mathcal{P}^{(0)}_2(y)}{y^2-z^2}  .
\end{gather}
Let us introduce $p_2(y) {=} \mathcal{P}^{(0)}_2(y)/y$, then we find
\begin{gather}
\mathcal{P}^{(0)}_1(z) =1+\frac{z}{\pi}\int\limits_0^\infty \frac{dy}{y} \Bigl [ p_2(y+z)-p_2(y-z)\Bigr ]  
\end{gather}
At $z{\to} 0$ we can expand the right hand side of the above equation in $z$, and find
\begin{gather}
\mathcal{P}^{(0)}_1(z) \simeq 1+\frac{2z^2}{\pi}\int\limits_0^\infty \frac{dy}{y} \Bigl [ p_2^\prime(y) + \frac{z^2}{6} p_2^{\prime\prime \prime}(y)\Bigr ] .
\end{gather}
We note that for $\eta{=}0$ the integral $\int_0^\infty dy\, p_2^\prime(y)/y{=}0$.  For $\eta{\neq} 0$ it is non-zero, and 
\begin{equation}
\mathcal{P}^{(0)}_1(z) \simeq 1+\frac{2z^2}{\pi}\int\limits_0^\infty \frac{dy}{y} [\mathcal{P}^{(0)}_2(y)/y]^\prime  .
\end{equation}

At $z\gg 1$ we shall use the following relation
\begin{gather}
\mathcal{P}^{(0)}_1(z) = \int\limits_0^\infty \frac{dy}{\pi y} \Bigl [\
\mathcal{P}^{(0)}_2(y+z)+\mathcal{P}^{(0)}_2(y-z)
\Bigr ] \notag \\
= \int\limits_0^\infty \frac{dy}{\pi y} 
\mathcal{P}^{(0)}_2(y+z)
-  \int\limits_0^z \frac{dy}{\pi y} \mathcal{P}^{(0)}_2(z-y)
\notag \\
+ \int\limits_z^\infty \frac{dy}{\pi y} 
\mathcal{P}^{(0)}_2(y-z) .
\label{eq:KK:P:rel}
\end{gather}
Now we estimate the integrals using the asymptotic \eqref{eq:P2:zL:new} of  $\mathcal{P}^{(0)}_2(y)$ at $y\gg 1$, 
\begin{gather}
\int\limits_0^\infty \frac{dy}{y} 
\mathcal{P}^{(0)}_2(y+z) \simeq C_\eta^{(\infty)} \int\limits_\delta^\infty
\frac{dy}{y(y+z)^{\gamma}} \notag \\
\simeq   C_\eta^{(\infty)} z^{-\gamma} \Bigl (\ln \frac{z}{\delta} -\gamma_E-\psi(\gamma)\Bigr ) ,
\notag \\
\int\limits_\delta^z \frac{dy}{y} 
\mathcal{P}^{(0)}_2(z-y) \simeq C_\eta^{(\infty)} \int\limits_\delta^z
\frac{dy}{y(z-y)^{\gamma}}  \notag \\
\simeq   C_\eta^{(\infty)} z^{-\gamma} \Bigl (\ln \frac{z}{\delta} -H(-\gamma)\Bigr ) ,
\notag \\
\int\limits_z^\infty \frac{dy}{y} 
\mathcal{P}^{(0)}_2(y-z) \simeq C_\eta^{(\infty)} \int\limits_z^\infty
\frac{dy}{y(y-z)^{\gamma}}  \notag \\
\simeq   C_\eta^{(\infty)} z^{-\gamma} \frac{\pi}{\sin(\pi \gamma)} ,
\label{eq:D15}
\end{gather}
where we remind $\gamma{=}(1{-}\eta)/(1{-}\eta/4){<}1$. Also $H(n){=}\sum_{i=1}^n1/i$ denotes the harmonic number. Therefore, we obtain at $z{\gg} 1$
\begin{gather}
\mathcal{P}^{(0)}_1(z) \propto 
z^{-\frac{1-\eta}{1-\eta/4}} .
\end{gather}
We note that due to the presence of $\ln z$ contributions in Eq. \eqref{eq:D15} it is not possible to determine the exact prefactor in the asymptotic expression for $\mathcal{P}^{(0)}_1(z)$.
Nevertheless, our approach guarantees to give the correct power-law behaviour of $\mathcal{P}^{(0)}_1(z)$ at $z{\gg}1$.

\section{Computation of asymptotic expression for 
the self energy \eqref{eq:ImS:00} \label{Appendix:4:Sigma:0}}

In this Appendix we present derivation of asymptotic expressions for the functions $\mathcal{F}_1^{(1)}(z)$ and $\mathcal{F}_2^{(1)}(z)$.
We use the approach similar to Appendix \ref{Appendix:4:PolOper:0}. At first, we introduce new variables $z{=}\omega/\omega_{\bm{k}},\;  x {=} q/k, \; a {=} y/k$, where $y {=} |\bm{k} {+} \bm{q}|$. Then after integration over $\Omega$ in Eq. \eqref{eq:ImSigma:Pert:1:R}, we find
\begin{gather}
    \operatorname{Im} \Sigma^{(1),R}_{\bm{k}}  (\omega) = \frac{\rho \omega \omega_{\bm{k}}}{6 A_{\eta}\pi^2}\int \frac{x^{2-2\eta}dxda \sin^3\varphi}{a^{3-\eta}}
    \notag \\
   \times \sum_{s=\pm} \mathcal{X}^{(0)}\left(\frac{z+s a^{2-\eta/2}}{x^{2-\eta/2}}\right) \frac{1}{\left(z + s a^{2-\eta/2}\right)} ,
        \label{eq::E1}
\end{gather}
where $\mathcal{X}^{(0)}(y){=}\mathcal{P}^{(0)}_2(y)/[|\mathcal{P}^{(0)}_1(y)|^2{+}|\mathcal{P}^{(0)}_2(y)|^2]$. 
Below we are not interested in numerical factors for reasons discussed in Appendix \ref{Appendix:3:Dyn:Pert}. Therefore, prefactors from now on will be omitted.

The integral in the form of Eq. \eqref{eq::E1} can be evaluated in different limits. For $z {\ll} 1$ we neglect $z$ under the integral signs and obtain
\begin{gather}
    \operatorname{Im} \Sigma^{(1),R}_{k} (\omega) =\frac{\rho \omega \omega_{\bm{k}}}{3A_{\eta}\pi^2}\int \frac{x^{2-2\eta}dxda \sin^3 \varphi}{a^{5-3\eta/2}} \mathcal{X}^{(0)}_2\left(\left(\frac{a}{x}\right)^{2-\eta/2}\right) .
\end{gather}
This integral converges. Therefore, for small $z$ imaginary part of the self energy behave according to Eq. \eqref{eq:F21:00:as}.

In order to analyze the limit $z{\gg}1$, we neglect $a$ in comparison with $z$ wherever it is possible. Then we obtain
\begin{gather}
    \operatorname{Im} \Sigma^{(1),R}_{k} (\omega) =\frac{\rho \omega \omega_{\bm{k}}}{3A_{\eta}\pi^2}\int\limits_{0}^{\infty}
    dx \frac{x^{2-2\eta}}{z}
    \mathcal{X}^{(0)}\left(\frac{z}{x^{2-\eta/2}}\right) \notag \\
        \times \int\limits_{|x-1|}^{x+1}\frac{da}{a^{3-\eta}}
        \left [1-\left(\frac{a^2-1-x^2}{2x}\right )^2\right ]^{3/2}
\end{gather}
To obtain the asymptotics of the above expression, we evaluate the integral over $a$ in two domains $x{\ll}1$ and $x{\gg}1$,
\begin{equation} \label{eq::asympt_int}
    \int\limits_{|x{-}1|}^{x{+}1}\frac{da}{a^{3{-}\eta}}
    \left [1{-}\left(\frac{a^2{-}1{-}x^2}{2x}\right )^2\right ]^{3/2} \approx   
    \begin{cases}
    \frac{3\pi x}{8},      & x \ll 1 ,\\
    \frac{3\pi}{8 x^{3-\eta}},  & x \gg 1 .
  \end{cases}
\end{equation}
Since we are only interested in the power dependence of the imaginary part of the self energy on frequency, $z$, we will integrate the asymptotic expression \eqref{eq::asympt_int} within the limits of the applicability of the approximation. Thus, we neglect the difference of the function from its asymptotics in a parametrically small region where this function has no singularities.
Then we find
\begin{gather}
    \operatorname{Im} \Sigma^{(1),R}_{k} (\omega) =\frac{\rho \omega \omega_{\bm{k}}}{3A_{\eta}\pi^2}\Bigl [\int\limits_{0}^{{\sim} 1} dx \frac{x^{3-2\eta}}{z} \mathcal{X}^{(0)}\left(\frac{z}{x^{2-\eta/2}}\right) \notag \\
    + \int\limits_{{\sim}1}^{\infty}
dx \frac{1}{z x^{1+\eta}}
\mathcal{X}^{(0)}\left(\frac{z}{x^{2-\eta/2}}\right)\Bigr ]
\end{gather}
Next, we use asymptotics of the polarization operator, cf. Eqs. \eqref{eq:ppp:01} and \eqref{eq:ReP:high:as}, to obtain
\begin{gather}
    \operatorname{Im} \Sigma^{(1),R}_{k} (\omega) \sim \rho \omega \omega_{\bm{k}} \frac{z^{\gamma}}{z}\Bigl [\int\limits_{0}^{{\sim}1}dx\, x 
    + \int\limits_{{\sim}1}^{\infty} \frac{dx}{x^{3{+}\eta/2}}
        \Bigr ]\notag \\
     \sim \rho \omega \omega_{\bm{k}} \bigl( \omega/\omega_k\bigr)^{\gamma-1} .
\end{gather}

In order to find the real part of the self energy and to derive Eq. \eqref{eq:F11:00:as}, one can employ the similar analysis as presented in Appendix \ref{Appendix:4:PolOper:0}.

\section{Computation of asymptotics for the exact polarization operator \eqref{eq:PolOp:main} \label{Appendix:6:PolOper:Exact}}

In this Appendix we 
present arguments for the scaling form \eqref{eq:PolOp:main} of the exact polarization operator and compute its asymptotic expressions. Let us consider the polarization operator computed as a bubble of the two exact Green's function, 
cf.  
Eq. \eqref{eq:pol:R:im}:
\begin{gather}
\im \Pi^{(b),R}_{\bm{q}}(\Omega) = 
\frac{2 T \Omega}{3}  \int \frac{d\omega}{2\pi}
\int \frac{d^2\bm{k}}{(2\pi)^2}
\frac{[\bm{k}\times \bm{q}]^4}{q^4} \frac{\im \mathcal{G}^{R}_{\bm{k}}(\omega)}{\omega}
\notag \\
\times \frac{\im \mathcal{G}^{R}_{\bm{k}+\bm{q}}(\omega+\Omega)}{\omega+\Omega} \ .
\label{eq::F1}
\end{gather}
For simplicity, we denote 
\begin{equation}
    \frac{\im \mathcal{G}^{R}_{\bm{k}}(\omega)}{\omega} = \frac{1}{\rho \omega_{\bm{k}}^3}\mathcal{A}\left(\frac{\omega}{\omega_{\bm{k}}}\right) ,
    \label{eq::F3}
\end{equation}
where 
\begin{gather}
    \mathcal{A}(z) = \frac{\mathcal{F}_2(z)}{\left[z^2 + \mathcal{F}_1(z)-1 \right]^2 + z^2\mathcal{F}^{2}_2(z)}.    
    \label{eq::F4}
\end{gather}
In order to make the integral dimensionless, we introduce new variables:
$\bm{y} {=} \bm{q} {+} \bm{k}$, $z {=} \Omega/\omega_{\bm{q}}$, $\tau {=} \omega/\omega_{\bm{q}}$, $k {=} x q$, $y {=} a q$, where $a{=}(1{+}x^2{+}2 x \cos\varphi)^{1/2}$. In terms of new variables the integral can be rewritten as
\begin{gather}
\im \Pi^{(b),R}_{\bm{q}} = \frac{2T z}{3 (2\pi)^3\varkappa^2q_*^{2\eta}q^{2-2\eta}} \int d\tau \int \frac{dx d\varphi \sin^4(\varphi)}{x^{1-3\eta/2}a^{6-3\eta/2}}\notag \\
\times \mathcal{A}\left(\frac{\tau}{x^{2-\eta/2}}\right) 
\mathcal{A}\left(\frac{z + \tau}{a^{2-\eta/2}}\right).
\end{gather}

First, we consider the case $z {\to} \infty$. In that limit we find
\begin{gather}
    \im \Pi^{(b),R}_{\bm{q}} = \frac{2T z}{3 (2\pi)^3\varkappa^2q_*^{2\eta}q^{2-2\eta}}  \int \frac{dx d\varphi \sin^4(\varphi)x^{1+\eta}}{a^{6-3\eta/2}}\notag \\
\times \mathcal{A}^{(1)}\left(\frac{z}{a^{2-\eta/2}}\right)\int \frac{d\tau}{x^{2-\eta/2}} \mathcal{A}^{(1)}\left(\frac{\tau}{x^{2-\eta/2}}\right) 
\end{gather}
The integral over $\tau$ converges and provides essentially a constant factor for the integral. 
Taking into account the fact, that the integral 
is dominated by  $a^{2{-}\eta/2}{\ll}z$ and $x {\gg} 1$, we substitute the asymptotic form of $\mathcal{A}(x)$ 
\begin{equation}
    \mathcal{A} (x) \sim x^{\gamma-5}, \quad x \gg 1.
    \label{eq::F7}
\end{equation}
In the region $x {\gg} 1$ we can substitute $a$ with $x$, thus separating integral over $\varphi$. Making these approximations, we find asymptotics of the imaginary part of the polarization operator as follows
\begin{gather}
    \im \Pi^{(b),R}_{\bm{q}} \sim \frac{T z}{ \varkappa^2q_*^{2\eta}q^{2-2\eta}}  \int_{1}^{z^{1/(2-\eta/2)}} \frac{dx x^{1+\eta}}{x^{6-3\eta/2}}\notag \\
\times \left(\frac{z}{x^{2-\eta/2}}\right)^{\gamma - 5} \sim  \frac{T}{ \varkappa^2q_*^{2\eta}q^{2-2\eta}} z^{-\gamma}
\end{gather}

In the limit of small frequencies, $z \to 0$, we find
\begin{gather}
\im \Pi^{(b),R}_{\bm{q}} = \frac{2T z}{3 (2\pi)^3\varkappa^2q_*^{2\eta}q^{2-2\eta}} \int d\tau \int \frac{dx d\varphi \sin^4(\varphi)}{x^{1-3\eta/2}a^{6-3\eta/2}}\notag \\
\times \mathcal{A}^{(1)}\left(\frac{\tau}{x^{2-\eta/2}}\right) 
\mathcal{A}^{(1)}\left(\frac{\tau}{a^{2-\eta/2}}\right) \sim z
\end{gather}
Therefore, using the exact Green's functions we reproduce exactly the same asymptotic expressions for the imaginary part of the polarization operator as we found in Appendix \ref{Appendix:4:PolOper:0}. Furthemore, due to the Kramers-Kronig relations, the scaling of asymptotic expression for the real part of the polarization operator is also the same as given in Appendix \ref{Appendix:4:PolOper:0}. We note that consideration of more complicated diagrams for the polarization operator does not change the scaling. 

In order to draw any conclusions we also need to check, whether self energy 
behaves in the way consistent with the form of the exact Green's function. This will be done in Appendix \ref{Appendix:7:Sigma:Exact}.

\section{Computation of asymptotics for exact self energy \eqref{eq:Green:main:0} \label{Appendix:7:Sigma:Exact}}

In this Appendix we present arguments for the scaling form \eqref{eq:Green:main:0} of the exact self energy and compute asymptotic expressions for the functions $\mathcal{F}_{1,2}(z)$.

Following the same analysis as in Appendix \ref{Appendix:6:PolOper:Exact}, we rewrite Eq. \eqref{eq:ImSigma:Pert:1:R}
as 
\begin{gather}
\Im \Sigma^{(1),R}_{\bm{k}}(\omega) {=} 
 {-} \frac{2T\omega}{3} \int \frac{d\Omega}{\pi}\!
\int \frac{d^2\bm{q}}{(2\pi)^2}
\frac{[\bm{k}\times \bm{q}]^4}{|\bm{k} + \bm{q}|^4}
\frac{\Im \mathcal{G}^{R}_{\bm{q}}(\Omega)}{\Omega(\omega+\Omega)}
\notag \\
\times  
\frac{\Im \Pi^{R}_{\bm{q}}(\Omega + \omega)}{|\Pi^{R}_{\bm{q}}(\Omega + \omega)|^2} .
\label{eq::G1}
\end{gather}
Here we use the exact polarization operator and exact Green's function. 
We use Eqs. \eqref{eq::F3} and \eqref{eq:PolOp:main} to rewrite the above expression as
\begin{gather}
 \Im \Sigma^{(1),R}_{\bm{k}}(\omega) {=} 
 {-} \frac{2\omega \varkappa^2 q_{*}^{2\eta}}{3\rho} \int \frac{d^2 \bm{q}}{(2\pi)^2}\frac{[\bm{k}\times \bm{q}]^4}{|\bm{k} + \bm{q}|^{2+2\eta}}
 \notag \\
 \times \int \frac{d\Omega}{\pi}\frac{\mathcal{P}_{2}\left(\frac{\omega + \Omega}{\omega_{\bm{k} + \bm{q}}}\right)}{\left |\mathcal{P}\left(\frac{\omega + \Omega}{\omega_{\bm{k} + \bm{q}}}\right)\right |^2(\omega + \Omega)}\frac{\mathcal{A}\left(\frac{\Omega}{\omega_{\bm{q}}}\right)}{\omega_{\bm{q}}^3}
 \label{eq:G2}
\end{gather}

In the limit $\omega \to \infty$ we use the fact,  that integral dominated by the region $\omega \gg |\Omega| \sim \omega_{\bm{q}} \sim \omega_{\bm{k}}$. Therefore, we can use the asymptotic expression for the polarization operator, found in Appendix \ref{Appendix:6:PolOper:Exact},
\begin{gather}
    \im \Sigma^{(1),R}_{\bm{k}}(\omega) {\sim} 
 {-} \frac{2\omega \varkappa^2 q_{*}^{2\eta}}{3\rho}\int \frac{d^2\bm{q}}{(2\pi)^2}\frac{[\bm{k}\times \bm{q}]^4}{|\bm{k} + \bm{q}|^{2+2\eta}}
 \notag \\
 \left(\frac{\omega}{\omega_{\bm{k} + \bm{q}}}\right)^{\gamma} \frac{1}{\omega \omega_{\bm{q}}^2}\int \frac{dy}{\pi}\mathcal{A}(y) ,
\end{gather}
where $y  = \Omega/\omega_{q}$. In virtue of Eq. \eqref{eq::F7}, the integral over $y$ converges. Introducing new dimensionless variables $\bm{q} = k\bm{r}$ and $\bm{k} = k \bm{n}$, we obtain
\begin{gather}
     \im \Sigma^{(1),R}_{\bm{k}}(\omega) \sim \rho \omega \omega_{\bm{k}} \left(\frac{\omega}{\omega_{\bm{k}}}\right)^{\gamma - 1}\int\frac{d^2\bm{r}}{(2\pi)^2}\frac{[\bm{n}\times\bm{r}]^4}{|\bm{n} + \bm{r}|^{4} r^{4-\eta}} \notag \\
     \times\int \frac{dy}{\pi}\mathcal{A}(y).
\end{gather}
Therefore, we reproduce Eq. \eqref{eq:F12:zz}.

In the opposite limit of small frequencies, $\omega \to 0$, we neglect $\omega$ under the integral sign in Eq. 
\eqref{eq:G2} and find
\begin{equation}
    \im \Sigma^{(1),R}_{\bm{k}}(\omega) \sim \rho \omega \omega_{\bm{k}} .
\end{equation}

Now using the Kramers-Kroning relations \eqref{eq:KK:0}, we find 
at $z{\gg}1$
\begin{gather}
\mathcal{F}_{1}(z)  =
{\rm p.v.} \int_0^\infty \frac{dz}{\pi} \frac{2 z^2 \mathcal{F}_2(x)}{x^2-z^2}
\simeq C^{(\infty)}_1 z^{\gamma} , \notag \\  C^{(\infty)}_1= C^{(\infty)}_2 \Phi_\gamma ,
\end{gather}
where 
\begin{equation}
    \Phi_\gamma  = {\rm p.v.} \int\limits_0^\infty \frac{dy}{\pi}\frac{2 y^\gamma}{y^2-1} = \int\limits_0^\infty \frac{dt}{\pi t} \Bigl [(1+t)^\gamma-|1-t|^\gamma\Bigr ] .
\end{equation}
Next applying the Kramers-Kroning relation again, we obtain
\begin{equation}
\mathcal{F}_{2}(z)  =
{\rm p.v.} \int_0^\infty \frac{dz}{\pi} \frac{2 \mathcal{F}_1(x)}{z^2-x^2} \simeq C_1^{\infty} z^{\gamma-1} \Phi_{\gamma-1} , 
\label{eq:F21:zz}
\end{equation}
In order Eq. \eqref{eq:F21:zz} to be mutually consistent with Eq. \eqref{eq:F12:zz}, the function $\Phi_\gamma$ has to satisfy the following relation 
$\Phi_\gamma \Phi_{\gamma-1}{=}-1$. It is indeed the case. Note that $\Phi_{-\gamma}=-\Phi_\gamma$.

The ongoing analysis has demonstrated that the inclusion of self energy correction in a self-consistent manner yields identical asymptotic outcomes for both the polarization operator and the self energy. This convergence indicates that universal scaling properties of the exact Green function 
are 
reproduced by 
SCSA-like diagrams. Extending this finding to encompass all correction diagrams requires recognizing a key observation: for every SCSA-like diagram, corresponding non-SCSA-like diagrams exist, characterized by the equivalent 
number of interaction ``wiggly'' lines, 
external momenta, and frequencies. This equivalence stems from the inherent limitation that interaction cannot transmit zero momentum.

Considering the power-law behavior of the self energy correction in terms of frequency and momentum, instilled by each SCSA-like diagram, the same behavior should be replicated by non-SCSA diagrams. Thus, the distinction lies mainly in numerical factors, with non-SCSA-like diagrams impacting only these specific coefficients.

\section{Computations within $1/d_c$ expansion\label{Appendix:8:1dc}}

In this Appendix we derive asymptotic results for the functions $\mathcal{P}_{1,2}(z)$ and $\mathcal{F}_{1,2}(z)$ within the $1/d_c$ expansion. In order to employ it, we consider 2D membrane embedded into $d_c+2$ dimensional space. Then $1/d_c$ can serve as the control parameter of the perturbative expansion in the screened interaction \cite{David1988}. In particular, the bending rigidity exponent $\eta$ is known to have the following expansion expansion \cite{Saykin2020}   
\begin{equation}
  \eta = \frac{2}{d_c} +  \frac{73-68\zeta(3)}{27 d_c^2} + \dots 
\end{equation}
Consequently, from Eq. \eqref{eq:def:gamma} we find  the following expansion for the exponent $\gamma$:
\begin{equation}
\gamma = 1 - \frac{3}{2 d_c} - \frac{25-17 \zeta(3)}{9 d_c^2} + \dots  
\label{eq:gamma:expansion}
\end{equation}

In Appendix \ref{Appendix:6:PolOper:Exact} we derived the exact form of the polarization operator. For it's imaginary part we obtained the asymptotic expressions \eqref{eq:PolOp:main:A1} and \eqref{eq:PolOp:main:A2}. In the limit $d_c{\to}\infty$ we can express $P_{1,2} (z)$ as a series expansion in terms of $1/d_c$. For example, for $z{\gg} 1$, we find
\begin{gather}
    P_{1} (z) = B_{1}^{(\infty)} z^{-\gamma} = (B_{1,0}^{(\infty)} + \frac{1}{d_c}B_{1,1}^{(\infty)} + \dots) \notag \\
    \times  \frac{1}{z} (1 + \frac{3}{2d_c}\ln |z| + \dots) 
    \label{eq:P1:appH}
\end{gather}
and similar expression for $P_{2}(z)$.

Since $1/d_c$ is a perturbation parameter, we can derive coefficients $B_{2,0}^{(\infty)}$ and $B_{2,0}^{(0)}$ by simply considering polarization operator, consisting of ``bare'' Green's functions. Thus, for imaginary part we obtain
\begin{gather}
\im \Pi^{(0),R}_{\bm{q}}(\Omega) = 
\frac{2 d_c T \Omega}{3}  \int \frac{d\omega}{2\pi}
\int \frac{d^2\bm{k}}{(2\pi)^2}
\frac{[\bm{k}\times \bm{q}]^4}{q^4} \frac{\im G^R_{\bm{k}}(\omega)}{\omega}
\notag \\
\times \frac{\im G^R_{\bm{k}+\bm{q}}(\omega+\Omega)}{\omega+\Omega} .
\end{gather}
where the imaginary part of ``bare'' Green's function is given by \eqref{eq::A3}. Performing integrals over frequency $\omega$ and momentum $\bm{k}$, we derive
\begin{equation}
\im \Pi^{(0),R}_{\bm{q}}(\Omega) = 
\frac{T}{\varkappa^2 q^2}
\mathcal{P}_2^{(0),0} \left(\frac{\Omega}{\omega_q} 
 \right ) . 
 \label{eq::H3}
 \end{equation}
Here the function $\mathcal{P}_2^{(0),0}(z)$ is odd, 
and for $z>0$ it is given as 
\begin{equation}
\mathcal{P}_2^{(0),0}(z)=\frac{d_c z}{96}\begin{cases}
\displaystyle 1,  & z< \frac{1}{2} ,\\
\displaystyle 1 + \frac{(1-2z)^2(z+1)}{2z^3},  & \frac{1}{2}\leqslant z <1 , \\
\displaystyle  \frac{(3z+1)}{2z^3}, & 1 \leqslant z .
\end{cases}
 \label{eq::H4}
\end{equation}
We use Eq. \eqref{eq::H4} in order to derive the expansion of the coefficients $B_{2}^{(0)}$ and  $B_{2}^{(\infty)}$ in powers $1/d_c$. In particular, we obtain
\begin{equation}
B_{2}^{(0)} = \frac{d_c}{96} + O(1), \quad 
B_{2}^{(\infty)}  = \frac{d_c}{64} + O(1).     
\end{equation}

In order to find the expansion for the coefficients $B_{1}^{(0)}$ and $B_{1}^{(\infty)}$, we need to compute the real part of the retarded polarization operator at finite frequency. With the help of the Kramers-Kronig relation, we find
\begin{gather}
\re \Pi^{(0),R}_{\bm{q}}(\Omega) = 
{\rm p.v.} \int\limits_{-\infty}^\infty \frac{d\omega}{\pi} \frac{\im \Pi^{(0),R}_{\bm{q}}(\omega)}{\omega-\Omega}  
\notag \\
= \frac{ T}{\varkappa^2 q^2} \mathcal{P}_1^{(0),0} \left(\frac{\Omega}{\omega_q} 
 \right ) ,
 \label{eq::H5}
\end{gather}
where the even function $\mathcal{P}_1^{(0),0}(z)$ is given explicitly as, 
\begin{align}
\mathcal{P}_1^{(0),0}(z) = & 
\frac{d_c}{192 \pi z^2} \Biggl [ (2 z+1)^2(z-1) \ln\bigl |1 + 2 z\bigr | + 8 z^2
 \notag \\
 & -(2 z-1)^2(z+1) \ln\bigl |1 - 2 z\bigr | 
 \notag \\
 &  +  6 z (1 - z^2)  \ln \left | \frac{1 + z}{1-z}\right | 
      \Biggr ] .
      \label{eq::H6}
\end{align}
Expanding $\mathcal{P}_1^{(0),0}(z)$ in series in powers of $z$, we obtain
\begin{equation}
\mathcal{P}_1^{(0),0} (0) = \frac{d_c}{16\pi} + O(1), \quad B_{1}^{(0)} = O\left(1\right) .
\end{equation}
Next, expanding at $z{\gg}1$ we derive $P_1^{(0),0}(z) {\sim} {-} d_c z^{-2} \ln z$. We see that it is a subleading contribution as one can see from Eq. \eqref{eq:P1:appH}. Therefore, in order to find $B_{1}^{(\infty)}$, straightforward usage of the result \eqref{eq::H6} is not possible. Instead we apply the relation \eqref{eq:Phi_gamma} and the following asymptotic expression for the function $\Phi_\gamma$
\begin{equation}
    \Phi_{1-\alpha} \simeq \frac{2}{\pi \alpha}, \quad \alpha \ll 1 .
    \label{eq::H7}
\end{equation}
Then we derive the following result 
\begin{equation}
B_{1}^{(\infty)} = - \frac{d_c^2}{96\pi} + O(d_c). 
\end{equation}

Similar procedure can be employed for the imaginary part of the self energy correction. Let us consider the lowest order (in $1/d_c$) correction  
\begin{gather}
\Im \Sigma^{(1),R}_{\bm{k}}(\omega) = 
 \frac{2T\omega}{3} \!\! \int \frac{d\Omega}{\pi} \!\!
\int \frac{d^2\bm{q}}{(2\pi)^2}
\frac{[\bm{k}{\times} \bm{q}]^4}{q^4}
\frac{\Im \Pi^{(0),R}_{\bm{q}}(\Omega)}{|\Pi^{(0),R}_{\bm{q}}(\Omega)|^2}
\notag \\
\times  
\frac{\Im G^R_{\bm{k}+\bm{q}}(\omega+\Omega)}{\Omega(\omega+\Omega)} .
\label{eq::H8}
\end{gather}
Substituting \eqref{eq::A4} into \eqref{eq::H8} and integrating over frequency, we obtain
\begin{gather}
    \Im \Sigma^{(1),R}_{\bm{k}}(\omega) = \frac{T\omega}{3\rho}\int\frac{d^2\bm{q}}{(2\pi)^2}\frac{\left[\bm{k}{\times}\bm{q}\right]^4}{q^4} \notag\\
    \times \sum_{s{=}{\pm}1} \frac{1}{\omega_{\bm{q} {+} \bm{k}}^{(0)2} (\omega {+} s\omega^{(0)}_{\bm{q} {+} \bm{k}})}
     \frac{\Im \Pi^{(0),R}_{\bm{q}}(\omega {+} s\omega^{(0)}_{\bm{q} {+} \bm{k}})}{|\Pi^{(0),R}_{\bm{q}}(\omega {+} s\omega^{(0)}_{\bm{q} {+} \bm{k}})|^2}  .
\end{gather}
In the limit $\omega {\ll} \omega_{\bm{k}}$ we can neglect the external frequency $\omega$ under the integral sign and obtain
\begin{gather}
    \Im \Sigma^{(1),R}_{\bm{k}}(\omega) = \frac{2T\omega}{3\rho}\int\frac{d^2\bm{q}}{(2\pi)^2}\frac{\left[\bm{k}\times\bm{q}\right]^4}{q^4 \omega_{\bm{q} {+} \bm{k}}^{(0)3}}
    \frac{\Im \Pi^{(0),R}_{\bm{q}}(\omega^{(0)}_{\bm{q} {+} \bm{k}})}{|\Pi^{(0),R}_{\bm{q}}(\omega^{(0)}_{\bm{q} {+} \bm{k}})|^2}  .
\end{gather}
Then with the help of Eqs. \eqref{eq::H4} and \eqref{eq::H6} the integral over $\bm{q}$ can be evaluated numerically. Hence, for $\omega{\ll}\omega_{\bm{k}}$ we obtain
\begin{gather}
    \Im \Sigma^{(1),R}_{\bm{k}}(\omega) \approx 1.57 \frac{\rho\omega\omega_{k}}{d_c}.
\end{gather}
Using the above asymptotic result, we find
\begin{equation}
\mathcal{F}_{2}(0)  = \frac{1.57}{d_c} + O\left(\frac{1}{d_c^2}\right) . 
\end{equation}

In order to determine asymptotics in the opposite limit, $\omega {\gg} \omega_{\bm{k}}$, we neglect $\omega_{\bm{q} {+} \bm{k}}$ in comparison with $\omega$ under the integral sign. Then we find
\begin{gather}
    \Im \Sigma^{(1),R}_{\bm{k}}(\omega) = \frac{2T}{3\rho}\int\frac{d^2\bm{q}}{(2\pi)^2}\frac{\left[\bm{k}{\times}\bm{q}\right]^4}{q^4 \omega_{\bm{q} {+} \bm{k}}^{(0)2}}\frac{\Im \Pi^{(0),R}_{\bm{q}}(\omega)}{|\Pi^{(0),R}_{\bm{q}}(\omega)|^2} .
\end{gather}

We proceed by substituting Eqs. \eqref{eq::H3} and \eqref{eq::H5} into the above expression. It is convenient to introduce new variables $\bm{k} {=} k\bm{n}$, $\bm{q} {=} k\bm{r}$, $z {=} \omega/\omega_{\bm{k}}$. Then we obtain
\begin{gather}
    \Im \Sigma^{(1),R}_{\bm{k}}(\omega) = \frac{2}{3 d_c}\rho\omega_{\bm{k}}^2\int\frac{d^2\bm{r}}{(2\pi)^2}\frac{\left[\bm{n}{\times}\bm{r}\right]^4}{r^2|\bm{r} {+} \bm{n}|^4} \notag \\
 \times\frac{\mathcal{P}^{(0),0}_2\left({z}/{r^2}\right)}{[\mathcal{P}^{(0),0}_1\left({z}/{r^2}\right)]^2+[\mathcal{P}^{(0),0}_2\left({z}/{r^2}\right)]^2} .
\end{gather}
In the limit $z{\gg}1$ we can use asymptotic expressions for $\mathcal{P}_2^{(0),0}(z)$ and $\mathcal{P}_1^{(0),0}(z)$ (see Eqs. \eqref{eq::H4} and \eqref{eq::H6}) since the integral is dominated by $r{\ll}\sqrt{z}$. Thus, we obtain
\begin{gather}
    \Im \Sigma^{(1),R}_{\bm{k}}(\omega) = \frac{32z}{3 d_c\pi^2}\rho\omega_{\bm{k}}^2\int\limits_{0}^{\infty} r dr \int\limits_{0}^{2\pi} \frac{d\varphi \, \sin^4 \varphi}{\left(r^2 {+} 1 {+} 2r\cos\varphi\right)^2}.
\end{gather}

The integral over angle $\varphi$ can be easily evaluated:
\begin{gather}
    \int\limits_{0}^{2\pi}d\varphi \frac{\sin^4\varphi}{(r^2 {+} 1 {+} 2r\cos\varphi)^2} = \frac{3\pi}{4}\begin{cases}
        1, &  r{\leqslant} 1 ,
 \\
 r^{-4}, &  r{>} 1 .
    \end{cases}
\end{gather} 

Finally, we obtain the asymptotic expression
\begin{gather}
    \Im \Sigma^{(1),R}_{\bm{k}}(\omega) \simeq \frac{8}{\pi d_c}\rho\omega\omega_{\bm{k}}, \quad \omega{\gg}\omega_{\bm{k}} .
\end{gather}
Using the above expression, we derive 
\begin{equation}
C_{2}^{(\infty)} = \frac{8}{\pi d_c} + O\left(\frac{1}{d_c^2}\right) .    
\end{equation}
In order to find the expansion of $C_{1}^{(\infty)}$ we use the relation \eqref{eq:Phi_gamma:2}. Then using Eq. \eqref{eq::H7}, we find
\begin{equation}
    C_1^{(\infty)} = \frac{32}{3\pi^2} + O\left (\frac{1}{d_c}\right ).
\end{equation}

\begin{figure}[t]
    \centering
    \includegraphics[width = 0.85\columnwidth]{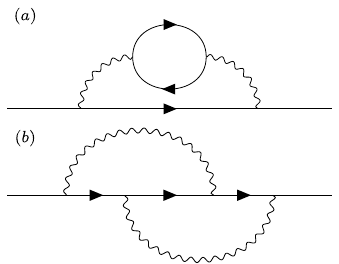}
    \caption{Diagrams for self energy in the second order of interaction with non-zero imaginary part.}
    \label{pic:NonAnomalous}
\end{figure}

\section{Calculation of the imaginary part of the self energy beyond the universal regime \label{App:I}}

We have shown, that in the region $q{\gg} q_*$ or $\Omega{\gg} \omega_*$ screening of the interaction is negligible in virtue of the condition $Y \Pi_{q}(\Omega) {\ll} 1$. Therefore, we could calculate decay rate for flexural phonons in this regime using perturbation theory.

In the first order in interaction the correction to the self energy is real. Thus, in order to calculate the attenuation, we need to study the second-order corrections. There are two diagrams in that order with non-zero imaginary part (see Fig. \ref{pic:NonAnomalous}). In this Appendix we will the present results for the diagram (a) only since the diagram (b) with crossed lines (b) is of the same or smaller magnitude.

We start from the following expression
\begin{gather}
    \im \Sigma_{\bm{k}}^{(a), R}(\omega)\sim T Y^2 \omega \int d\Omega \int d^2 \bm{q} \frac{\left[\bm{k} \times \bm{q}\right]^4}{q^4} \frac{\im \Pi^{R}_{\bm{q}}(\Omega)}{\Omega}\notag \\ \times\frac{\im G^R_{\bm{q}+\bm{k}}(\omega+\Omega)}{\omega+\Omega}  .
\end{gather}

We first consider the case $k{\gg}q_*$ (regions II$_b$ and III in Fig. \ref{fig:UniversalRegime}). Making the integral dimensionless, we obtain
\begin{gather}
\im \Sigma_{\bm{k}}^{(a), R}(\omega) = \left(\frac{q_*}{k}\right)^4 \rho \omega \omega_{\bm{k}} f_2^{(a)} (z),
\end{gather}
where $z {=} \omega / \omega_{\bm{k}}$ and $\omega_{\bm{k}} {=} Dk^2$. After straightforward calculations we obtain asymptotics
\begin{gather}
    f_2^{(a)} (z) \sim 
    \begin{cases}
    \textrm{const}, & \quad |z| \ll 1 , \\
    1/z^2, & \quad |z| \gg 1 .
    \end{cases}
    \label{eq::I3}
\end{gather}
Therefore, in the regime $k {\gg} q_*$, we find
\begin{equation}
    \im \Sigma_{\bm{k}}^{(a), R}(\omega) \sim \left(\frac{\omega_*}{\max\{\omega, \omega_{\bm{k}}\}}\right)^2 \rho \omega \omega_{\bm{k}}.
\end{equation}
This form suggests that the obtained correction is small in virtue of a small parameter $q_*/k {\ll} 1$ that controls the perturbation theory.

In the region $k{\ll} q_*$ and $\omega {\gg}\omega_*$ (region II$_a$ in Fig. \ref{fig:UniversalRegime}), we need to account for renormalization of the bending rigidity. Thus we obtain
\begin{gather}
    \im \Sigma_{\bm{k}}^{(a), R}(\omega) \sim \frac{T^2 Y^2 \omega}{\varkappa^2}\int d^2 \bm{q} \frac{\left[\bm{k} \times \bm{q}\right]^4
    }{|\bm{k} + \bm{q}|^6} \frac{1}{\rho \omega_{\bm{q}}^2 }\notag \\
    \times \sum_{s{\pm}1} \mathcal{P}_2\left(\frac{\omega+s\omega_{\bm{q}}}{D|\bm{k} + \bm{q}|^2}\right)\frac{1}{\omega+s\omega_{\bm{q}}}.
\end{gather}
The integral over momentum is dominated by $q{\sim}k$, therefore, we have to use  $\omega_{\bm{q}} {=} Dq^{2-\eta/2}q_{*}^{\eta/2}$. Evaluating the integral over $q$, we find
\begin{gather}
     \im \Sigma_{\bm{k}}^{(a), R}(\omega) \sim 
     \rho \omega_{*}^2 \left(\frac{k}{q_*}\right)^{\eta}\frac{Dk^2}{\omega}  .
\end{gather}

A special care is needed for calculation of the imaginary part of the self energy at the mass shell $\omega {=} \omega_{\bm{k}} {\gg} \omega_{*}$. As one can see, the following contribution
\begin{gather}
    \im \Sigma_{\bm{k}}^{(a), R}(\omega) \sim \frac{T^2 Y^2 \omega}{\varkappa^2}\int d^2 \bm{q} \frac{\left[\bm{k} \times \bm{q}\right]^4
    }{q^6} \notag \\
        \times \mathcal{P}_2\left(\frac{\omega_{\bm{k} +\bm{k} }-\omega_{\bm{k}}}{\omega_{\bm{q}}}\right)\frac{1}{\rho \omega_{\bm{q}}^2 }\frac{1}{\omega_{\bm{k} + \bm{q}}- \omega_{\bm{k}}}
\end{gather}
diverges due to singularity at $q{\to}0$. To fix this problem, one has to work with the full RPA screened interaction. Then we obtain
\begin{gather}
    \im \Sigma_{\bm{k}}^{(a), R}(\omega) \sim T \omega_{\bm{k}} \int d^2 \bm{q} \frac{Y^2 \im \Pi^R_{\bm{q}}\left(\omega_{\bm{k}+\bm{q}} - \omega_{\bm{k}}\right)}{| 1+3Y \Pi^R_{\bm{q}}\left(\omega_{\bm{k}+\bm{q}} - \omega_{\bm{k}}\right)/2|^2} \notag \\
    \times\frac{1}{\rho \omega_{\bm{k}+\bm{q}}^2}\frac{\left[\bm{k} \times \bm{q}\right]^4}{q^4} \frac{1}{\omega_{\bm{k}} - \omega_{\bm{k}+\bm{q}}}.
\end{gather}
The integral over $q$ is now dominated by $q {\sim} q_{*}$ rather than $q{=}0$. This justifies the usage of the full form of the RPA screened interaction. Using asymptotics for the imaginary part of the polarization operator, Eq. \eqref{eq:PolOp:main:A2}, we obtain final result
\begin{gather}
    \im \Sigma_{\bm{k}}^{(a), R}(\omega) \sim \rho \omega_{*}^2 \frac{k}{q_{*}}.
\end{gather}
This result is valid for $|\omega {-} D k^2| {\ll} \omega_*$ since in that region integral over $q$ is also dominated by $q{\sim}q_*$ and the same approximations have to be employed.

	
\bibliography{biblio-elasticity-f}

\end{document}